\documentclass[12pt]{article}
\usepackage{latexsym}
\usepackage{amsmath,amsfonts}
\usepackage{times}
\allowdisplaybreaks[4]

\catcode`\@=11

\newcount\hour
\newcount\minute
\newtoks\amorpm \hour=\time\divide\hour by 60\minute
=\time{\multiply\hour by 60 \global\advance\minute by-\hour}
\edef\standardtime{{\ifnum\hour<12 \global\amorpm={am}%
        \else\global\amorpm={pm}\advance\hour by-12 \fi
        \ifnum\hour=0 \hour=12 \fi
        \number\hour:\ifnum\minute<10
        0\fi\number\minute\the\amorpm}}
\edef\militarytime{\number\hour:\ifnum\minute<10 0\fi\number\minute}

\def\draftlabel#1{{\@bsphack\if@filesw {\let\thepage\relax
   \xdef\@gtempa{\write\@auxout{\string
      \newlabel{#1}{{\@currentlabel}{\thepage}}}}}\@gtempa
   \if@nobreak \ifvmode\nobreak\fi\fi\fi\@esphack}
        \gdef\@eqnlabel{#1}}
\def\@eqnlabel{}
\def\@vacuum{}
\def\marginnote#1{}
\def\draftmarginnote#1{\marginpar{\raggedright\scriptsize\tt#1}}
\def\draft{
        \pagestyle{plain}
        \overfullrule=2pt
        \oddsidemargin -.5truein
        \def\@oddhead{\sl \phantom{\today\quad\militarytime} \hfil
        \smash{\Large\sl DRAFT} \hfil \today\quad\militarytime}
        \let\@evenhead\@oddhead
        \let\label=\draftlabel
        \let\marginnote=\draftmarginnote
        \def\ps@empty{\let\@mkboth\@gobbletwo
        \def\@oddfoot{\hfil \smash{\Large\sl DRAFT} \hfil}
        \let\@evenfoot\@oddhead}
        \def\@eqnnum{(\theequation)\rlap{\kern\marginparsep\tt\@eqnlabel}%
        \global\let\@eqnlabel\@vacuum}  }

\newcommand{\rf}[1]{(\ref{#1})}
\renewcommand{\theequation}{\thesection.\arabic{equation}}
\renewcommand{\thefootnote}{\fnsymbol{footnote}}
\newcommand{\newsection}{   
\setcounter{equation}{0}\section}

\def\appendix#1{\addtocounter{section}{1}\setcounter{equation}{0}
\renewcommand{\thesection}{\Alph{section}}
\section*{Appendix \thesection\protect\indent \parbox[t]{11.15cm}{#1}}
\addcontentsline{toc}{section}{Appendix \thesection\ \ \ #1}}

\def\be{\begin{equation}}
\def\ee{\end{equation}}
\def\beq{\begin{eqnarray}}
\def\eeq{\end{eqnarray}}

\def\parline{\,\partial\kern -0.55em /\,\,}

\def\half{{\frac{1}{2}}}

\def\DD{{\cal D}}
\def\FF{{\cal F}}
\def\LL{{\cal L}}

\def\RR{{\cal R}}

\def\deltabf{{\boldsymbol{\delta}}}

\def\smK{{\scriptscriptstyle K}}
\def\smR{{\scriptscriptstyle R}}

\def\smzero{{\scriptscriptstyle (0)}}

\def\smzero{{\scriptscriptstyle (0)}}

\def\Dwh{\widehat{D}}
\def\Rwh{\widehat{R}}
\def\Gwh{\widehat{G}}

\def\omegawh{\widehat{\omega}}

\def\deltabf{{\boldsymbol{\delta}}}

\def\smzero{{\scriptscriptstyle 0}}
\def\smminone{{\scriptscriptstyle -1}}
\def\smmintwo{{\scriptscriptstyle -2}}
\def\smminthree{{\scriptscriptstyle -3}}

\def\flat{{\rm flat}}

\def\lin{{\rm lin}}
\def\scal{{\rm scal}}
\def\weyl{{\rm weyl}}

\def\Gbf{{\bf G}}
\def\Rbf{{\bf R}}
\def\gbf{{\bf g}}
\def\nbf{{\bf n}}

\begin{document}


\begin{flushright}
FIAN-TD-2010-14 \hspace{1.6cm}{}~\\
arXiv: 1012.2079 [hep-th]\quad{}~\\
\end{flushright}

\vspace{1cm}

\begin{center}

{\Large \bf 6d conformal gravity }

\vspace{2.5cm}

R.R. Metsaev\footnote{ E-mail: metsaev@lpi.ru }

\vspace{1cm}

{\it Department of Theoretical Physics, P.N. Lebedev Physical
Institute, \\ Leninsky prospect 53,  Moscow 119991, Russia }

\vspace{3.5cm}

{\bf Abstract}

\end{center}

In the framework of ordinary-derivative approach, conformal gravity in
space-time of dimension  six is studied. The field content, in addition to
conformal graviton field, includes  two auxiliary rank-2 symmetric tensor
fields, two Stueckelberg vector fields and one Stueckelberg scalar field.
Gauge invariant Lagrangian with conventional kinetic terms and the
corresponding gauge transformations are obtained. One of the rank-2 tensor
fields and the scalar field have canonical conformal dimension. With respect
to these fields, the Lagrangian contains, in addition to other terms, a cubic
potential. Gauging away the Stueckelberg fields and excluding the auxiliary
fields via equations of motion, the higher-derivative Lagrangian of $6d$
conformal gravity is obtained. The higher-derivative Lagrangian involves
quadratic and cubic curvature terms. This higher-derivative Lagrangian
coincides with the simplest Weyl invariant density discussed in the earlier
literature. Generalization of de Donder gauge conditions to $6d$ conformal
fields is also obtained.

\newpage
\renewcommand{\thefootnote}{\arabic{footnote}}
\setcounter{footnote}{0}

\section{Introduction}

 In view of their aesthetic features, conformal fields  have attracted
considerable interest for a long period of time (see
Ref.\cite{Fradkin:1985am,Vasiliev:2009ck,Segal:2002gd}). In space-time of
dimension $d\geq 4$, conformal fields can be separated into two groups:
fundamental  fields and shadow fields.
A field
having Lorentz algebra spin $s$ and conformal dimension $\Delta = s +d-2$ is
referred to as fundamental field, while a  field having Lorentz algebra spin $s$
and dual conformal dimension $\Delta = 2-s$ is  referred to as
shadow field%
\footnote{ Lorentz algebra label $s$ is used for description of the totally
symmetric fields. To discuss so called mixed-symmetry fields one needs to
involve more labels of the Lorentz algebra. Discussion of conformal
mixed-symmetry fields may be found in
Refs.\cite{Vasiliev:2009ck,Shaynkman:2004vu}.}.
In this paper we deal only with shadow fields which  will be referred
to simply as conformal fields in what follows.

The conformal fields are used, among other things, to discuss conformally
invariant Lagrangians (see e.g.
\cite{Fradkin:1985am,Vasiliev:2009ck,Segal:2002gd}). With the exception of
some particular cases, Lagrangian formulation of the conformal fields involve
higher derivatives and non-conventional kinetic terms. We note also that the
higher derivative terms  hide propagating degrees of freedom (d.o.f) of
conformal fields. In Refs.\cite{Metsaev:2008ba,Metsaev:2007fq} we developed
ordinary (not higher-) derivative, gauge invariant  Lagrangian formulation
for {\it free} conformal fields. That is  to say that our Lagrangians for
bosonic fields do not involve higher than second order terms in derivatives
and have  conventional kinetic terms.

In this paper we discuss $6d$ conformal gravity using the framework of
ordinary-derivative approach developed in Ref.\cite{Metsaev:2007fq}. The
purpose of this paper is to develop an ordinary-derivative, gauge invariant,
and Lagrangian formulation for {\it interacting} fields of $6d$ conformal
gravity%
\footnote{ Ordinary-derivative approach to interacting $4d$ conformal gravity
was discussed in Ref.\cite{Metsaev:2007fq}.}.
Our approach to the interacting conformal $6d$ gravity can be summarized as
follows.

i) We introduce additional field degrees of freedom, i.e., we extend the
space of fields entering the standard $6d$ conformal gravity. In addition to
conformal graviton field, our field content involves two rank-2 symmetric
tensor fields, two vector fields and one scalar field.
All additional fields are supplemented by appropriate gauge symmetries%
\footnote{ To realize those additional gauge symmetries we adopt the approach
of Refs.\cite{Metsaev:2008ba}-\cite{Metsaev:2006zy} which turns out to be the
most useful for our purposes.}.
We note that the vector fields and the scalar field turn out to be
Stueckelberg fields, i.e. they are somewhat similar to the ones used in the
gauge invariant approach to massive fields.

ii) Our Lagrangian for interacting fields of the $6d$ conformal gravity does
not contain  higher than second order terms in derivatives. To second order
in fields, two-derivative contributions to the Lagrangian take the form of
the standard kinetic terms of the scalar, vector, and tensor fields. Two
derivative contributions appear also in the interaction vertices.

iii) Gauge transformations of fields $6d$ conformal gravity do not involve
higher than first order terms in derivatives. Interacting independent
one-derivative contributions to the gauge transformations take the form of
the standard gauge transformations of the vector and tensor fields.

iv) The gauge symmetries of our Lagrangian make it possible to match  our
approach with  the higher-derivative one, i.e., by an appropriate gauge
fixing of the Stueckelberg fields and by solving some constraints we obtain
the higher-derivative formulation of the $6d$ conformal gravity. This implies
that our approach retain propagating d.o.f of the higher-derivative $6d$
conformal gravity theory, i.e., our approach is equivalent to the
higher-derivative one, at least at classical level.

As is well known, the Stueckelberg approach turned out to be successful for
the study of theories involving massive fields (see e.g.
Ref.\cite{Siegel:1985tw}). In fact, all covariant formulations of
string theories are realized by using Stueckelberg gauge symmetries.
Therefore we expect that use of the Stueckelberg fields for the studying
conformal fields might be useful for developing new interesting formulations
of the $6d$ conformal theory.

The rest of the paper  is organized as follows.

Sec. \ref{sec-03} is devoted to the discussion of free spin-2 conformal field
in $6d$ flat space. In Sec. \ref{sec-03-sub01} we start with brief review of
the higher-derivative formulation of free $6d$ conformal gravity. After this,
in Sec. \ref{sec03-02}, we review ordinary-derivative formulation of free
$6d$ conformal gravity. We discuss various representation for the gauge
invariant Lagrangian. We review gauge symmetries of the Lagrangian and
realization of global conformal algebra symmetries on the space of gauge
fields.

In Sec. \ref{man02-sec-04} we describe the ordinary-derivative formulation of
interacting theory of $6d$ conformal gravity. We discuss ordinary-derivative
gauge invariant Lagrangian and its gauge symmetries.

In Sec. \ref{sec-04-a1}, we show how the higher-derivative Lagrangian of
interacting $6d$ conformal gravity is obtained from our ordinary-derivative
Lagrangian.

Section \ref{conl-sec-01} suggests directions for future research.

In Appendix A, we summarize our conventions and the notation. In Appendix B
we present some details of the derivation of gauge invariant Lagrangian and
the corresponding gauge transformations.

\newsection{ Free spin-2 conformal field in $6d$ flat space}\label{sec-03}

To make contact with studies in earlier literature we start with presentation
of the standard, i.e. higher-derivative, formulation for the spin-2 conformal
field propagating in $6d$ flat space.  In the literature, such field is often
referred to as conformal Weyl graviton.

\subsection{ Higher-derivative formulation of  spin-2 conformal field}\label{sec-03-sub01}

To discuss higher-derivative and gauge invariant formulation of spin-2
conformal field one uses rank-2 the Lorentz algebra $so(5,1)$ tensor field
$\phi^{ab}$ having conformal dimension $\Delta_{\phi^{ab}}=0$. The field
$\phi^{ab}$ is symmetric, $\phi^{ab}=\phi^{ba}$, and traceful $\phi^{aa} \ne
0$. Higher-derivative Lagrangian for the field $\phi^{ab}$ is given by
\be \label{Lconfie2lag01}  \LL= \frac{1}{3}C_\lin^{abce} \Box C_\lin^{abce}\,, \ee
where $C_\lin^{abce}$ is the linearized Weyl tensor. Using representation of
the Weyl tensor in terms of curvatures
\beq
C^{abce} &=& R^{abce} - \frac{1}{4}(\eta^{ac}R^{be} -
\eta^{bc}R^{ae} + \eta^{be}R^{ac} - \eta^{ae}R^{bc})
\nonumber\\
& + & \frac{1}{20}(\eta^{ac} \eta^{be} - \eta^{ae} \eta^{bc}) R\,,
\eeq
and the Gauss-Bonnet relation
\be R_\lin^{abce} \Box R_\lin^{abce} - 4 R_\lin^{ab} \Box R_\lin^{ab} +
R_\lin\Box R_\lin = 0 \quad (\hbox{ up to total derivative})\,, \ee
we obtain the representation for Lagrangian \rf{Lconfie2lag01} in terms of linearized Ricci
curvatures,
\be \label{2m04122010-02}
\LL = R_\lin^{ab} \Box R_\lin^{ab} - \frac{3}{10} R_\lin^2 \,,\ee
which is also useful for certain purposes. Using explicit representation of the
Ricci curvatures in terms of the field $\phi^{ab}$
\beq
\label{2m04122010-06} && \hspace{1cm} R_\lin^{ab} =
\frac{1}{2}\left(-\Box\phi^{ab} + \partial^a \partial^c\phi^{bc} +\partial^b
\partial^c\phi^{ac} - \partial^a\partial^b \phi^{cc} \right)\,,
\\[5pt]
\label{2m04122010-07} && \hspace{1cm} R_\lin  = \partial^a\partial^b
\phi^{ab} - \Box \phi^{aa}\,,
\eeq
leads to other well known form of the Lagrangian
\be \label{Lconfie2lag01n01} \LL = \frac{1}{4} \phi^{ab} \Box^3 P^{ab\, ce}
\phi^{ce} \,, \ee
where we use the notation as in \cite{Fradkin:1985am}:
\be
P^{ab\, ce} \equiv \half(\pi^{ac}\pi^{be} +  \pi^{ae}\pi^{bc})  -
\frac{1}{5} \pi^{ab}\pi^{ce} \,,
\qquad \pi^{ab} \equiv \eta^{ab}   - \frac{\partial^a \partial^b}{\Box} \,.
\ee
Lagrangian \rf{Lconfie2lag01} is invariant under linearized
diffeomorphism and Weyl gauge transformations
\beq
&& \label{secspin2con10} \delta \phi^{ab} = \partial^a \xi^b +
\partial^b \xi^a + \eta^{ab} \xi\,,
\eeq
where $\xi^a$ and $\xi$ are the respective diffeomorphism and Weyl gauge
transformation parameters.

We now discuss on-shell d.o.f of $6d$ conformal gravity. To this end we use
fields transforming in irreps of $so(4)$ algebra. One can prove that on-shell
d.o.f are described by three rank-2 {\it traceless} symmetric tensor fields
$\phi_{k'}^{ij}$, two vector fields $\phi_{k'}^i$, and one scalar field
$\phi_0$:
\footnote{ Fields \rf{spin2DoF01} are related to non-unitary representation
of conformal algebra $so(6,2)$. Discussion of unitary representations of
the conformal algebra that are relevant for elementary particles may be found
e.g. in Refs.\cite{Siegel:1988gd},\cite{Metsaev:1995jp}.}
\beq
& \phi_\smmintwo^{ij}\qquad \phi_0^{ij}\qquad \phi_2^{ij} &
\nonumber\\[5pt]
\label{spin2DoF01} & \phi_\smminone^i\qquad \phi_1^i &
\\[5pt]
& \phi_0 &
\nonumber
\eeq
$i,j=1,\ldots,4$ (for details see Appendix B in Ref.\cite{Metsaev:2007fq}).
Total number of on-shell d.o.f shown in \rf{spin2DoF01} is given by
\be \label{spin2DoF04}
\nbf = 36 \,.
\ee
We note that this $\nbf$ is decomposed in a sum of d.o.f for fields given in
\rf{spin2DoF01} as%
\footnote{ Total d.o.f given in \rf{spin2DoF04} was found in
Ref.\cite{Fradkin:1985am}. Decomposition of $\nbf$ \rf{spin2DoF05} into irreps
of the $so(4)$ algebra was carried out in Ref.\cite{Metsaev:2007fq} (see
Appendix B in Ref.\cite{Metsaev:2007fq}). Light-cone gauge approach used in
Ref.\cite{Metsaev:2007fq} provides easy possibility to decompose the total
$\nbf$ into irreps of $so(4)$ algebra. Discussion of other methods for
counting propagating d.o.f of higher-derivative theories may be found in
Refs.\cite{Lee:1982cp,Buchbinder:1987vp}.}:

\beq \label{spin2DoF05}
&& \nbf = \sum_{k'=0,\pm 2} \nbf (\phi_{k'}^{ij}) + \sum_{k'=\pm 1}
\nbf(\phi_{k'}^i) + \nbf(\phi_0)\,,
\\[5pt]
&& \hspace{2cm} \nbf(\phi_{k'}^{ij}) = 9, \hspace{2.3cm} k'= 0,\pm 2\,;
\\[5pt]
&& \hspace{2cm}  \nbf(\phi_{k'}^i) = 4, \hspace{2.3cm} k'=\pm 1\,;
\\[5pt]
&&  \hspace{2cm} \nbf(\phi_0) = 1\,.
\eeq

\subsection{ Ordinary-derivative formulation of spin-2 conformal
field}\label{sec03-02}

We now review the ordinary-derivative formulation of the spin-2 conformal
field in $6d$ flat space developed in Ref.\cite{Metsaev:2007fq}. In addition
to results in Ref.\cite{Metsaev:2007fq}, we discuss also two new
representations for gauge invariant Lagrangian. One of the new
representations turns out to be convenient for the generalization to theory
of interacting spin-2 conformal field. Also we present our results for de
Donder like gauge conditions which have not been discussed in the earlier
literature.

{\bf Field content}. To discuss ordinary-derivative and gauge invariant
formulation of the spin-2 conformal field in $6d$ flat space we use three
rank-2 tensor fields $\phi_{k'}^{ab}$, two vector fields $\phi_{k'}^a$, and
one scalar field $\phi_{0}$:
\beq \label{covspin2DoF01}
& \phi_{-2}^{ab}\qquad \phi_0^{ab}\qquad \phi_2^{ab} &
\nonumber\\[5pt]
& \phi_{-1}^a\qquad \phi_1^a &
\\[5pt]
& \phi_0 &
\nonumber
\eeq
The fields $\phi_{k'}^{ab}$, $\phi_{k'}^a$ and $\phi_{0}$ are the respective
rank-2 tensor, vector, and scalar fields of the Lorentz algebra $so(5,1)$.
Note that the tensor fields $\phi_{k'}^{ab}$ are symmetric, $\phi_{k'}^{ab}=
\phi_{k'}^{ba}$, and traceful, $\phi_{k'}^{aa}\ne 0$. Fields in
\rf{covspin2DoF01} have the conformal dimensions
\beq
\label{Pdelspi2def01} && \Delta_{\phi_{k'}^{ab}} = 2 + k'\,, \qquad k' = 0,\pm 2\,,
\nonumber\\[3pt]
&& \Delta_{\phi_{k'}^a} = 2 + k'\,, \qquad k'=\pm 1\,,
\\[3pt]
&& \Delta_{\phi_{0}} = 2\,.
\nonumber
\eeq

{\bf Gauge invariant Lagrangian}. We discuss three representations for
Lagrangian in turn.

{\it 1st representation for the Lagrangian}. This representation found in
Ref.\cite{Metsaev:2007fq} is given by
\beq \label{2man29112010-04}
\LL & = & \frac{1}{2} \phi_2^{ab} (E_{_{EH}}\phi_{-2})^{ab} + \frac{1}{4}
\phi_0^{ab} (E_{_{EH}}\phi_0)^{ab} +  \phi_{1}^a (E_{_{Max}} \phi_{-1})^a +
\half \phi_{0} \Box \phi_0
\nonumber\\[3pt]
& + &  \phi_{1}^a \partial^b \chi_0^{ab}
+ \phi_{-1}^a \partial^b \chi_2^{ba}
- \half \chi_0^{ab} \phi_2^{ab}\,,
\\[5pt]
\label{2m23122010-01} &&  \hspace{2cm} \chi_0^{ab} \equiv \phi_0^{ab} -\eta^{ab}\phi_0^{cc} - u
\eta^{ab}\phi_0\,,
\\[5pt]
\label{2m23122010-02} &&  \hspace{2cm} \chi_2^{ab} \equiv \phi_2^{ab} -\eta^{ab}\phi_2^{cc} \,,
\\[5pt]
\label{2m01122010-03} &&  \hspace{3cm} u \equiv \sqrt{5/2}\,,
\eeq
where $E_{_{EH}}$ and $E_{_{Max}}$ are the respective second-order
Einstein-Hilbert and Maxwell operators,
\beq \label{2m04122010-09}
&& (E_{_{EH}}\phi)^{ab} = \Box \phi^{ab} -\partial^a\partial^c\phi^{cb} -
\partial^b\partial^c\phi^{ca}  + \partial^a \partial^b \phi^{cc}
+  \eta^{ab}(\partial^c\partial^e\phi^{ce} - \Box \phi^{cc}) \,,\qquad
\\[5pt]
&& (E_{_{Max}}\phi)^a = \Box \phi^a  -\partial^a \partial^b\phi^b \,.
\eeq
Thus, we see that two-derivative contributions to Lagrangian
\rf{2man29112010-04} takes the form of standard second-order kinetic terms
for the respective rank-2 tensor fields, vector fields and scalar field. Note
also that besides the two-derivative contributions, the Lagrangian involves
one-derivative contributions and derivative-independent mass-like
contributions.

{\it 2nd representation for the Lagrangian}. The second representation has
not been discussed in the earlier literature. For the reader convenience, we
discuss this representation because it allows us to introduce de Donder like
gauge
conditions for $6d$ conformal gravity%
\footnote{ De Donder gauges turn out to be useful for study of various
dynamical systems. Recent discussion of the {\it standard} de Donder-Feynman
gauge for massless fields may be found in
Refs.\cite{Manvelyan:2008ks,Fotopoulos:2009iw,Guttenberg:2008qe}.
Applications of  modified de Donder gauges for massless and massive $AdS$
fields \cite{Metsaev:2008ks} to studying the $AdS/CFT$ correspondence may found
in  Ref.\cite{Metsaev:2008fs}.}.
This is to say that Lagrangian \rf{2man29112010-04} can be represented as (up
to total derivative)
\beq
\LL & = & \frac{1}{2} \phi_2^{ab} \Box \phi_{-2}^{ab} - \frac{1}{4}
\phi_2^{aa} \Box \phi_{-2}^{bb} + \frac{1}{4} \phi_0^{ab} \Box \phi_0^{ab} -
\frac{1}{8} \phi_0^{aa} \Box \phi_0^{bb} + \phi_{1}^a \Box \phi_{-1}^a +
\half \phi_{0} \Box \phi_0 \qquad\quad
\nonumber\\[3pt]
& +  & C_{-1}^a C_3^a + \half C_1^a C_1^a + C_0 C_2
\\[3pt]
& - &  \half \phi_2^{ab} \phi_0^{ab} + \frac{1}{4} \phi_2^{aa}
\phi_0^{bb} - \half \phi_1^a \phi_1^a\,,
\nonumber
\eeq
where quantities $C_{k'}^a$, $C_{k'}$, which we refer to as conformal de
Donder divergences, are given by
\beq
&& C_{-1}^a  = \partial^b \phi_{-2}^{ab} - \half \partial^a \phi_{-2}^{bb} +
\phi_{-1}^a\,,
\\[5pt]
&& C_1^a = \partial^b \phi_0^{ab} - \half \partial^a \phi_0^{bb} +
\phi_1^a\,,
\\[5pt]
&& C_3^a = \partial^b \phi_2^{ab} - \half \partial^a \phi_2^{bb}\,,
\\[5pt]
&& C_0 =  \partial^a \phi_{-1}^a + \half \phi_0^{aa} + u\phi_0\,,
\\[5pt]
&& C_2 =  \partial^a \phi_1^a + \half \phi_2^{aa}\,.
\eeq
We note that it is the conformal de Donder divergencies that define de Donder like
gauge conditions for our conformal $6d$ fields,
\beq
&& C_{k'}^a = 0 \,,\qquad k'= ,-1,1,3\,,
\nonumber\\[-7pt]
&& \hspace{6cm} \hbox{conformal de Donder gauge conditions.}
\\[-7pt]
&& C_{k'} = 0 \,,\qquad k'= 0,2\,.
\nonumber
\eeq

The fields described by Lagrangian (2.24) are related to non-unitary
representation of the conformal algebra. Fields with $k'=0$ have kinetic
terms with correct signs. The remaining fields with $k'\ne 0$ can be
collected into the vector and tensor doublets, $\phi_{-1}^a$,  $\phi_1^a$ and
$\phi_{-2}^{ab}$,  $\phi_2^{ab}$. We note then that vector (tensor) doublet
describes one vector (tensor) field with wrong sign of kinetic term and one
vector (tensor) field with correct sign of the  kinetic term.

{\it 3rd representation for the Lagrangian}. Finally, we discuss
representation for free Lagrangian \rf{2man29112010-04} which turns to be
most adapted for generalization to interacting $6d$ conformal gravity. This
is to say that Lagrangian \rf{2man29112010-04} can be represented as (up to
total derivative)
\beq
\label{2m04122010-01a1} && \hspace{3cm} \LL = \sum_{a=1}^6 \LL_a \,,
\\
\label{2m04122010-01a2} && \LL_1  = -\phi_2^{ab} \Gwh_\lin^{(ab)}\,,
\\[5pt]
\label{2m04122010-01a3} &&  \LL_2 = -\frac{1}{4} \partial^c\phi_0^{ab} \partial^c
\phi_0^{ab} + \frac{1}{8} \partial^c\phi_0^{aa}\partial^c \phi_0^{bb} + \half
C_1^a C_1^a \,,
\\[5pt]
\label{2m04122010-01a4} &&  \LL_3 = - \half
F^{ab}(\phi_\smminone)F^{ab}(\phi_1)\,,
\\
\label{2m04122010-01a5} && \LL_4 = - \half \partial^a \phi_0
\partial^a\phi_0\,,
\\[5pt]
\label{2m04122010-01a6} && \LL_5 = \phi_1^a \partial^b \chi_0^{ab}\,,
\\[5pt]
\label{2m04122010-01a7} && \LL_6 = -\half \phi_2^{ab}\chi_0^{ab}\,,
\\[5pt]
\label{2m04122010-01a8} && \hspace{1cm} \Gwh_\lin^{(ab)} = G_\lin^{ab} +
\half (\partial^a\phi_{-1}^b +
\partial^b\phi_{-1}^a)  - \eta^{ab}\partial^c\phi_{-1}^c\,,
\\[5pt]
\label{2m04122010-08} && \hspace{1cm} G_\lin^{ab} = R_\lin^{ab}(\phi_\smmintwo) - \half
\eta^{ab} R_\lin(\phi_\smmintwo) \,,
\\
&& \hspace{1cm} C_1^a \equiv \partial^b \phi_0^{ab} - \half \partial^a \phi_0^{bb}\,,
\\
&& \hspace{1cm} F^{ab}(\phi_{k'}) \equiv \partial^a \phi_{k'}^b -
\partial^b\phi_{k'}^a\,,\qquad k' = \pm 1\,,
\eeq
where $\chi_0^{ab}$ is defined in \rf{2m23122010-01}. The linearized Ricci
curvatures for the field $\phi_\smmintwo^{ab}$ in \rf{2m04122010-08} are
obtained by substituting the field $\phi_\smmintwo^{ab}$  in the respective
expressions \rf{2m04122010-06} and \rf{2m04122010-07}. Note that linearized
Einstein tensor $G_\lin^{ab}$ \rf{2m04122010-08} can be represented by using
operator $E_{_{EH}}$ \rf{2m04122010-09} as
\be \label{2m17122010-01}
G_\lin^{ab} = -\half (E_{_{EH}}\phi_\smmintwo)^{ab} \,.
\ee
Also, we note that shifted linearized Einstein tensor $\Gwh^{ab}$
\rf{2m04122010-01a8} can be expressed in terms of the corresponding shifted
linearized Ricci curvatures
\be  \label{2m17122010-02} \Gwh_\lin^{ab} = \Rwh_\lin^{ab} - \half \eta^{ab}
\Rwh_\lin\,,\ee
where the shifted linearized curvatures are defined by relations
\beq  \label{2m17122010-03}
&& \Rwh_\lin^{abce}  =  R_\lin^{abce}  + \eta^{ac} \varphi_\lin^{be} -
\eta^{bc} \varphi_\lin^{ae} + \eta^{be} \varphi_\lin^{ac} - \eta^{ae}
\varphi_\lin^{bc}\,,
\\
&& R_\lin^{abce} = \half (-\partial^a\partial^c \phi_\smmintwo^{be} + \partial^b\partial^c \phi_\smmintwo^{ae} - \partial^b\partial^e \phi_\smmintwo^{ac} + \partial^a\partial^e \phi_\smmintwo^{bc})\,,
\\
\label{2m17122010-04} && \Rwh_\lin^{ab}  =  R_\lin^{ab}  +
4\varphi_\lin^{ab} +\eta^{ab} \varphi_\lin^{cc}\,,
\\[5pt]
\label{2m17122010-05} && \Rwh_\lin  =  R_\lin  + 10\varphi_\lin^{cc}\,,
\\[5pt]
\label{2m17122010-06} && \varphi_\lin^{ab} = q \partial^a \phi_\smminone^b\,, \qquad q =\frac{1}{4}\,,
\\[5pt]
\label{2m17122010-07} && \Rwh^{ab} = \Rwh^{cacb}\,,\qquad \Rwh = \Rwh^{aa}\,.
\eeq

{\bf Gauge transformations}. We now discuss gauge symmetries of Lagrangian
\rf{2man29112010-04}. To this end we introduce the gauge transformation
parameters,

\beq \label{2m08122010-08}
& \xi_\smminthree^a\qquad \xi_\smminone^a\qquad \xi_1^a &
\nonumber\\[-7pt]
&&
\\[-7pt]
& \xi_\smmintwo\qquad \xi_0&
\nonumber
\eeq
Conformal dimensions of the gauge transformation parameters are given by
\beq
&& \Delta_{\xi_{k'}^a}= 2 + k' \,, \qquad k ' = -3,-1,1\,,
\nonumber\\[-7pt]
&&
\\[-7pt]
&& \Delta_{\xi_{k'}} = 2 + k'\,, \qquad k' = -2,0\,.
\nonumber
\eeq
The gauge transformation parameters $\xi_{k'}^a$ and $\xi_{k'}$ are the
respective vector and scalar fields of the Lorentz algebra $so(5,1)$. The
Lagrangian is invariant under the gauge transformations
\beq \label{2m01122010-05}
&& \delta \phi_\smmintwo^{ab} = \partial^a \xi_\smminthree^b + \partial^b
\xi_\smminthree^a + \half \eta^{ab}\xi_\smmintwo\,,
\\[5pt]
\label{2m01122010-06} && \delta \phi_0^{ab} = \partial^a \xi_\smminone^b +
\partial^b \xi_\smminone^a + \half \eta^{ab}\xi_0\,,
\\[5pt]
\label{2m01122010-07} && \delta \phi_2^{ab} = \partial^a \xi_1^b + \partial^b
\xi_1^a\,,
\\[5pt]
\label{2man29112010-01} && \delta \phi_\smminone^a  = \partial^a
\xi_\smmintwo - \xi_\smminone^a\,,
\\[5pt]
\label{2man29112010-02}  && \delta \phi_1^a = \partial^a \xi_0 - \xi_1^a\,,
\\[5pt]
\label{2man29112010-03}  && \delta \phi_0 = - u \xi_0\,,
\eeq
where $u$ is given in \rf{2m01122010-03}.

{\bf Realization of conformal algebra symmetries}. In $6d$ space-time, the
conformal algebra $so(6,2)$ referred to the basis of Lorentz algebra
$so(5,1)$ consists of translation generators $P^a$, conformal boost
generators $K^a$, dilatation generator $D$ and generators of the Lorentz
algebra $so(5,1)$ denoted by $J^{ab}$. We assume the following normalization
for commutators of the conformal algebra%
\footnote{ Note that in our approach only $so(5,1)$ symmetries are realized
manifestly. The $so(6,2)$ symmetries could be realized manifestly by using
ambient space approaches (see e.g.
\cite{Preitschopf:1998ei,Bekaert:2009fg,Bonezzi:2010jr})}:
\beq
\label{ppkk}
&& {}[D,P^a]=-P^a\,, \hspace{2cm}  {}[P^a,J^{bc}]=\eta^{ab}P^c -\eta^{ac}P^b
\,,
\nonumber\\
\label{dppkk} && [D,K^a]=K^a\,, \hspace{2.2cm} [K^a,J^{bc}]=\eta^{ab}K^c -
\eta^{ac}K^b\,,
\nonumber\\[-7pt]
&&
\\[-7pt]
\label{pkjj} && \hspace{2.5cm} {}[P^a,K^b]=\eta^{ab}D-J^{ab}\,,
\nonumber\\
&& \hspace{2.5cm} [J^{ab},J^{ce}]=\eta^{bc}J^{ae}+3\hbox{ terms} \,.
\nonumber
\eeq
Let $\phi$ denotes field propagating in the flat space-time. Let Lagrangian
for the free field $\phi$ be conformal invariant. This implies, that
Lagrangian is invariant with respect to transformation (invariance of the
Lagrangian is assumed to be up to total derivative)
\be \label{man-12112009-03} \delta_{\hat{G}} \phi  = \hat{G} \phi \,, \ee
where a realization of the conformal algebra generators $\hat{G}$ in terms of
differential operators acting on $\phi$ takes the form
\beq
\label{conalggenlis01} && P^a = \partial^a \,,
\\[3pt]
\label{conalggenlis02} && J^{ab} = x^a\partial^b -  x^b\partial^a + M^{ab}\,,
\\[3pt]
\label{conalggenlis03} && D = x^a\partial^a  + \Delta\,,
\\[3pt]
\label{conalggenlis04} && K^a = K_{\Delta,M}^a + R^a\,,
\\[3pt]
\label{conalggenlis05} &&\qquad  K_{\Delta,M}^a \equiv
-\frac{1}{2}x^bx^b\partial^a + x^a D + M^{ab}x^b\,.
\eeq
In \rf{conalggenlis02}-\rf{conalggenlis04}, $\Delta$ is operator of conformal
dimension, $M^{ab}$ is spin operator of the Lorentz algebra. Action of
$M^{ab}$ on fields of the Lorentz algebra is well known and for rank-2
tensor, vector, and scalar fields considered in this paper is given by
\beq
&& M^{ab}\phi^{ce} = \eta^{ae}\phi^{cb} +\eta^{ac}\phi^{be} -
(a\leftrightarrow b)\,,
\nonumber\\
&& M^{ab}\phi^c = \eta^{ac}\phi^b - (a\leftrightarrow b)\,,
\\
&& M^{ab}\phi = 0\,.
\nonumber
\eeq
Relation \rf{conalggenlis04} implies that conformal boost transformations can
be presented as
\be \label{2m22122010-10} \delta_{K^a} \phi = \delta_{K_{\Delta,M}^a}\phi +
\delta_{R^a} \phi \,.\ee
Explicit representation for the action of operator $K_{\Delta,M}^a$
\rf{conalggenlis05} is easily obtained from the relations above-given. This
is to say that the rank-2 tensor, vector, and scalar fields considered in this paper
transform as
\beq \label{2m22122010-11}
&&  \delta_{K_{\Delta,M}^a}\phi_{k'}^{bc}= K_{\Delta(\phi_{k'})}^a
\phi_{k'}^{bc} + M^{abf} \phi_{k'}^{fc} + M^{acf} \phi_{k'}^{bf}\,,
\hspace{1.5cm} k'=0,\pm 2\,,\qquad\quad
\nonumber\\[5pt]
&& \delta_{K_{\Delta,M}^a}\phi^b = K_{\Delta(\phi_k')}^a \phi_{k'}^b +
M^{abf} \phi_{k'}^f\,,\hspace{3.8cm} k'= \pm 1 \,,
\\[5pt]
&& \delta_{K_{\Delta,M}^a}\phi_0 = K_{\Delta(\phi_0)}^a \phi_0\,,
\nonumber
\\[5pt]
\label{man10112009-04} && \hspace{3cm}  K_{\Delta}^a \equiv -\half x^bx^b
\partial^a + x^a (x\partial + \Delta) \,,
\\
&& \hspace{3cm}  M^{abc} \equiv \eta^{ab}x ^c -\eta^{ac} x^b \,. \eeq
Thus, all that remains is to find explicit representation for operator $R^a$
in \rf{conalggenlis04}. The operator $R^a$ depends on the derivative
$\partial^a$ and does not depend on the space-time coordinates $x^a$,
$[P^a,R^b]=0$. In the standard formulation of the conformal fields, the
operator $R^a$ is equal to zero, while in the ordinary-derivative approach,
we discuss in this paper, the operator $R^a$ is non-trivial. This implies
that, in the framework of ordinary-derivative approach, the complete
description of the conformal fields requires finding not only gauge invariant
Lagrangian but also the operator $R^a$. Realization of the operator $R^a$ on
a space of gauge fields \rf{covspin2DoF01} is given by
\beq
\label{24122009-01} && \delta_{R^a} \phi_{-2}^{bc} = 0\,,
\\[5pt]
\label{24122009-02} && \delta_{R^a} \phi_0^{bc} =   - 2(\eta^{ab}\phi_{-1}^c
+ \eta^{ac}\phi_{-1}^b) + \eta^{bc}\phi_{-1}^a - 4\partial^a
\phi_{-2}^{bc}\,,
\\[5pt]
\label{24122009-03} && \delta_{R^a} \phi_2^{bc} =  - 4(\eta^{ab}\phi_1^c +
\eta^{ac}\phi_1^b) + 2\eta^{bc}\phi_1^a - 4\partial^a \phi_0^{bc}\,,
\\[5pt]
\label{24122009-04} && \delta_{R^a} \phi_{-1}^b  =  4 \phi_{-2}^{ab}\,,
\\[5pt]
\label{24122009-05} && \delta_{R^a} \phi_1^b  =  2\phi_0^{ab} - 2 u \eta^{ab}
\phi_0 - 2 \partial^a \phi_{-1}^b\,,
\\[5pt]
\label{24122009-06} && \delta_{R^a} \phi_0  =  2 u \phi_{-1}^a \,.
\eeq
From \rf{24122009-01}-\rf{24122009-06}, we see the operator $R^a$ maps the
gauge field with conformal dimension $\Delta$ into the ones having conformal
dimension less than $\Delta$. This is to say that the realization of the
operator $R^a$ given in \rf{24122009-01}-\rf{24122009-06} can schematically
be represented as%
\footnote{ Realization of the operator $R^a$ on space of on-shell fields  can
be obtained by using group theoretical methods, while the realization of
$R^a$ on space of gauge fields requires the use of the gauge invariant
approach.}
\beq
&&  \phi_2^{ab} \stackrel{R}{\longrightarrow} \phi_1^a  \oplus
\partial \phi_0^{ab}\,,
\hspace{1.8cm}
\phi_0^{ab} \stackrel{R}{\longrightarrow} \phi_{-1}^a  \oplus
\partial \phi_{-2}^{ab}\,,
\quad\qquad
\phi_{-2}^{ab} \stackrel{R}{\longrightarrow} 0 \,,
\\[7pt]
&& \phi_1^a \stackrel{R}{\longrightarrow} \phi_0^{ab} \oplus \phi_0\oplus
\partial \phi_{-1}^a\,,\qquad \phi_0\stackrel{R}{\longrightarrow} \phi_{-1}^a\,,
\hspace{2.8cm} \phi_{-1}^a \stackrel{R}{\longrightarrow} \phi_{-2}^{ab}\,.
\qquad\eeq

{\bf Interrelation of the ordinary-derivative and the higher-derivative
approaches}. From \rf{2man29112010-01}-\rf{2man29112010-03} , we see that
both vector fields $\phi_\smminone^a$, $\phi_1^a$ and the scalar field
$\phi_0$ transforms as Stueckelberg (Goldstone) fields under the gauge
transformations, i.e. these fields can be gauged away by using the gauge
symmetries. Gauging away the vector fields and the scalar field,
\be \label{2man29112010-05} \phi_{\pm 1}^a = 0 \,,\qquad \phi_0 = 0 \,,\ee
we see that our Lagrangian \rf{2m04122010-01a1} takes the simplified form
\be \label{2man29112010-06}
\LL  = -\phi_2^{ab} G_\lin^{ab}
-\frac{1}{4} \partial^c\phi_0^{ab} \partial^c
\phi_0^{ab} + \frac{1}{8} \partial^c\phi_0^{aa}\partial^c \phi_0^{bb} + \half
C_1^a C_1^a
-\half \phi_2^{ab}\chi_0^{ab}\,.
\ee
Now using equations of motion for the rank-2 tensor field $\phi_2^{ab}$
obtained from Lagrangian \rf{2man29112010-06} we find the equation
\be \label{2man29112010-06a1} \phi_0^{ab} - \eta^{ab}\phi_0^{cc} = -
2G_\lin^{ab}\,, \ee
which has the obvious solution
\be \label{2man29112010-06a2} \bar\phi_0^{ab} = - 2 R_\lin^{ab} +
\frac{1}{5}\eta^{ab} R_\lin\,, \ee
where the linearized Ricci curvatures are obtained by substituting the field
$\phi_\smmintwo^{ab}$ in \rf{2m04122010-06},\rf{2m04122010-07}. Plugging
solution $\bar\phi_0^{ab}$ \rf{2man29112010-06a2} into Lagrangian
\rf{2man29112010-06} we obtain the higher-derivative Lagrangian given in
\rf{2m04122010-02}. Thus we see that our ordinary-derivative approach is
equivalent to the standard one and our field $\phi_\smmintwo^{ab}$ is
identified with excitation of the conformal graviton field $\phi^{ab}$ in
Sec. \ref{sec-03-sub01}.

\newsection{ Interacting $6d$ conformal gravity } \label{man02-sec-04}

We begin our discussion of interacting theory of $6d$ conformal gravity with
the description of a field content. Field content of the interacting theory
is simply obtained by promoting the Minkowski space free fields
\rf{covspin2DoF01} to the fields in curved space-time described by metric
tensor field $g_{\mu\nu}$. As usually, this metric tensor field is considered
to be conformal graviton field. As we have already said, the field
$\phi_{-2}^{ab}$ describes excitation of the conformal graviton, i.e., in the
interacting theory, the field $\phi_\smmintwo^{ab}$ is related to the metric
tensor field $g_{\mu\nu}$. Also note that, instead of metric-like approach to
conformal gravity, we prefer to use frame-like approach, i.e., we use
vielbein field $e_\mu^a$, $g_{\mu\nu} = e_\mu^a e_\nu^a$ and fields carrying
tangent-flat indices, $\phi^a$, $\phi^{ab}$, which are related to fields
carrying base manifold indices $\phi^\mu$, $\phi^{\mu\nu}$, by the standard
relations $\phi^a = e_\mu^a\phi^\mu$, $\phi^{ab}
=e_\mu^ae_\nu^b\phi^{\mu\nu}$ (for details of our notation see Appendix A).
Also, following commonly used nomenclature, we use notation $b^a$ in place of
the field $\phi_{-1}^a$. To summarize, the field content we use to develop
the ordinary-derivative approach to the interacting $6d$ conformal gravity is
given by%
\footnote{ The symmetric and antisymmetric parts of the gauge field
associated with the conformal boosts are related to the field $\phi_0^{ab}$
and the field strength $F^{ab}(b)$ (see \rf{2m04122010-03}) respectively.
Also, we note that the parameter $\xi_{-1}^a$ (see \rf{2m01122010-02}) is
related to conformal boosts gauge transformation parameter.}
\beq \label{2m04122010-05}
& e_\mu^a \qquad \phi_0^{ab}\qquad \phi_2^{ab} &
\nonumber\\[5pt]
& b^a\qquad \phi_1^a &
\\[5pt]
& \phi_0 &
\nonumber
\eeq
For field $\phi$ having Weyl dimension $\Delta_\phi^w$, we define local Weyl
transformations in the usual way,
\be \label{2m07122010-01} \delta \phi  =  \Delta_\phi^w \sigma \phi\,,\ee
where $\sigma$ is Weyl gauge transformation parameter. Using this convention, the
Weyl dimensions of the fields are given by%
\footnote{ In Ref.\cite{Fradkin:1985am}, conformal dimension is referred to
as canonical dimension.},
\beq \label{2m07122010-02}
&& \Delta_{e_\mu^a}^w = -1 \,, \qquad  \Delta_{\phi_{k'}^{ab}}^w = 2 + k'\,,
\qquad k' = 0,2\,,
\nonumber\\[-7pt]
&&
\\[-7pt]
&& \Delta_{\phi_1^a}^w = 3 \,, \qquad \ \ \Delta_{\phi_0}^w = 2 \,.
\nonumber
\eeq
Gauge transformation of the compensator field $b^a$ involves gradient term
(see below), but for constant $\sigma$ the field $b^a$ transforms as in
\rf{2m07122010-01} with $\Delta_{b^a}^w = 1$. With this convention for the
Weyl dimension of the field $b^a$, we note that conformal dimensions of
fields not carrying base manifold indices, $\phi_0^{ab}$, $\phi_2^{ab}$,
$b^a$, $\phi_1^a$, $\phi_0$, \rf{Pdelspi2def01} are equal to their Weyl
dimensions \rf{2m07122010-02}.

We now discuss gauge invariant Lagrangian for interacting fields
\rf{2m04122010-05}. The Lagrangian we find is given by
\beq
\label{2m04122010-01} && \hspace{5cm} \LL = \sum_{a=1}^8 \LL_a \,,
\\
\label{22112010-01} && e^{-1} \LL_1 = -\phi_2^{ab} \Gwh^{(ab)} \,,
\\[5pt]
\label{22112010-02} && e^{-1}\LL_2 = - \frac{1}{4}\DD^a \phi_0^{bc}\DD^a
\phi_0^{bc} + \frac{1}{8}\DD^a \phi_0^{bb}\DD^a \phi_0^{cc} + \half C_1^a C_1^a
\nonumber\\[5pt]
&& \hspace{1.3cm}  - \,\, \half \Rwh^{cabe} \phi_0^{ab} \phi_0^{ce} + \half
\Rwh^{ab} \phi_0^{ac}\phi_0^{cb} -  \half \Rwh^{ab} \phi_0^{ab} \phi_0^{cc} +
(\frac{1}{8} \phi_0^{aa}\phi_0^{bb} - \frac{1}{4} \phi_0^{ab}\phi_0^{ab})
\Rwh\,,\qquad
\\[5pt]
\label{22112010-03} && e^{-1} \LL_3 = - \half \FF^{ab}(b)\FF^{ab}(\phi_1)\,,
\\[5pt]
\label{22112010-04} && e^{-1} \LL_4 = - \half \DD^a\phi_0\DD^a\phi_0\,,
\\[5pt]
\label{22112010-05} && e^{-1} \LL_5 = \phi_1^a \DD^b \chi_0^{ab}\,,
\\[5pt]
\label{22112010-06} && e^{-1} \LL_6 = -\half \phi_2^{ab}\chi_0^{ab}\,,
\\[5pt]
\label{22112010-07} && e^{-1} \LL_7 = \frac{1}{4}\phi_0^{ab} T^{ab}
- \frac{u}{8}\phi_0 F^{ab}F^{ab}\,,
\\[5pt]
\label{22112010-09} &&  e^{-1} \LL_8 =
\frac{1}{4}\phi_0^{ab}\phi_0^{bc}\phi_0^{ca}
-\frac{5}{16}\phi_0^{ab}\phi_0^{ab}\phi_0^{cc} + \frac{1}{16}(\phi_0^{aa})^3
\nonumber\\[5pt]
&& \hspace{1cm}  - \,\, \frac{u}{8}\phi_0^{ab}\phi_0^{ab}\phi_0 -
\frac{3}{16}\phi_0^{aa}\phi_0^2 - \frac{3}{16u}\phi_0^3\,,
\\[5pt]
&& \hspace{1cm} e \equiv \det e_\mu^a\,,
\\[5pt]
\label{22112010-13} && \hspace{1cm} \Gwh^{(ab)} \equiv  G^{ab} + \half (D^a
b^b + D^b b^a) + \frac{1}{4} b^a b^b - \eta^{ab}( D^cb^c - \frac{3}{8}
b^cb^c)\,,
\\[5pt]
\label{22112010-13a1} && \hspace{1cm} G^{ab} \equiv R^{ab} -\half \eta^{ab}
R\,,
\\[5pt]
\label{22112010-14} && \hspace{1cm} C_1^a \equiv \DD^b \phi_0^{ab} - \half
\DD^a\phi_0^{bb}\,,
\\[5pt]
&&  \hspace{1cm} \chi_0^{ab} \equiv \phi_0^{ab} -\eta^{ab}\phi_0^{cc} - u \eta^{ab}\phi_0\,,
\\[5pt]
\label{22112010-11} && \hspace{1cm} T^{ab} \equiv F^{ac}F^{bc} -
\frac{1}{4}\eta^{ab}F^{ce}F^{ce}\,,
\\[5pt]
\label{22112010-11a1} && \hspace{1cm} F^{ab} \equiv D^a b^b - D^b b^a\,,
\eeq
where $u$ is defined in \rf{2m01122010-03}. Complete description of our
notation may be found in Appendix A. Here we mention the most important
notation.

\noindent {\bf a}) For rank-$s$ field $\phi^{b_1\ldots b_s}$ having Weyl dimension $\Delta_\phi^w$, covariant derivative $\DD^a$ is defined to be
\be \label{2m08122010-01} \DD^a\phi^{b_1\ldots b_s}  = \Dwh^a\phi^{b_1\ldots
b_s} + \Delta_{\phi}^w q b^a \phi^{b_1\ldots b_s}\,,
\quad\qquad  q = \frac{1}{4}\,, \ee
where $\Dwh^a$ is covariant derivative with shifted Lorentz connection
$\omegawh_\mu^{ab}$,
\beq
&& \Dwh^a \phi^b = e^{\mu a}\partial_\mu \phi^b + \omegawh^{abc} \phi^c
\,,\hspace{2.8cm} \omegawh^{abc} = e^{\mu a}\omegawh_\mu^{bc}\,,
\\[5pt]
&& D^a \phi^b = e^{\mu a}\partial_\mu \phi^b + \omega^{abc} \phi^c
\,,\hspace{2.8cm} \omega^{abc} = e^{\mu a}\omega_\mu^{bc}\,,
\\[5pt]
&& \omegawh^{abc} = \omega^{abc} + q (\eta^{ac} b^b - \eta^{ab} b^c)\,,
\eeq
while $D^a$ is a covariant derivative with the standard Lorentz connection
$\omega_\mu^{bc}(e)$.

\noindent {\bf b}) Field strength $\FF^{ab}(\phi)$ for vector field $\phi^a$ is defined to be
\be \label{2m04122010-03} \FF^{ab}(\phi) \equiv \DD^a\phi^b - \DD^b \phi^a\,.
\ee
Note that for the compensator field $b^a$ the field strength $\FF^{ab}$
becomes the standard one \rf{22112010-11a1},
\be \label{2m04122010-04} \FF^{ab}(b) = F^{ab}\,. \ee

\noindent {\bf c}) Curvature $\Rwh^{abce}$ is defined for the shifted
connection $\omegawh_\mu^{ab}$ as
\beq \label{2m04122010-04a1}
&& \Rwh_{\mu\nu}{}^{ab} = \partial_\mu \omegawh_\nu^{ab} -\partial_\nu
\omegawh_\mu^{ab} + \omegawh_\mu^{ac}\omegawh_\nu^{cb} -
\omegawh_\nu^{ac}\omegawh_\mu^{cb}\,,
\\[5pt]
&&  \Rwh^{abce} = e^{\mu a} e^{\nu b} \Rwh_{\mu\nu}{}^{ce}\,.
\eeq
Ricci curvatures $R^{ab}$, $R$ are defined as $R^{ab} = R^{cacb}$, $R =
R^{aa}$, where $R^{abce} =e^{\mu a} e^{\nu b}R_{\mu\nu}{}^{ce}$ and
$R_{\mu\nu}{}^{ab}$ is curvature for the standard Lorentz connection
$\omega_\mu^{ab}(e)$. We note that the shifted Einstein tensor is defined in
a usual way
\be \label{2m17122010-08}  \Gwh^{ab} = \Rwh^{ab} - \half
\eta^{ab}\Rwh\,,\qquad \Rwh^{ab} = \Rwh^{cacb}\,,\qquad \Rwh = \Rwh^{aa}\,.
\ee

A few remarks are in order.

\noindent {\bf i}) Contributions to interacting Lagrangian \rf{2m04122010-01}
denoted by $\LL_1$, $\LL_3$, $\LL_4$, $\LL_5$ are obtained by
covariantization of the flat derivative, $\partial^a \rightarrow \DD^a$, and
the linearized Einstein tensor, $G_\lin^{ab}\rightarrow \Gwh^{ab}$, in the
respective contributions $\LL_1$, $\LL_3$, $\LL_4$, $\LL_5$ to flat
Lagrangian \rf{2m04122010-01a1}. Note also that contribution $\LL_6$
\rf{2m04122010-01a7} in the flat Lagrangian is promoted to the interacting
Lagrangian without any changes (see \rf{22112010-06}).

\noindent {\bf ii}) Comparing $\LL_2$ \rf{22112010-02} that enters
interacting Lagrangian \rf{2m04122010-01} and the respective  $\LL_2$
\rf{2m04122010-01a3} that enters flat Lagrangian \rf{2m04122010-01a1}, we see
that $\LL_2$ in  \rf{22112010-02} involves the additional contributions of first
order in the shifted curvatures and second order in the field $\phi_0^{ab}$.
We note the $\LL_2|_{b^a=0}$ part of $\LL_2$ \rf{22112010-02} is obtained
simply by expanding the Einstein-Hilbert Lagrangian $\sqrt{\gbf}R(\gbf)$ with
$\gbf_{\mu\nu} = g_{\mu\nu} + \phi_{0,\mu\nu}$ as a power series in the field
$\phi_{0,\mu\nu}$ and taking terms of the second order in the field
$\phi_{0,\mu\nu}$, where $\phi_{0,\mu\nu} = e_{\mu a}e_{\nu b}\phi_0^{ab}$.
So, we see that the tensor field $\phi_0^{ab}$ looks like a excitation of the
graviton field.

\noindent {\bf iii}) Interesting contribution which is absent in flat
Lagrangian \rf{2m04122010-01a1} and involved in interacting Lagrangian \rf{2m04122010-01} is governed by
the contribution denoted by $\LL_7$ \rf{22112010-07}. From  \rf{22112010-07}, we
see that the tensor field $\phi_0^{ab}$ is coupled to energy-momentum tensor
of the compensator field $b^a$, i.e., we see that the tensor field
$\phi_0^{ab}$ again exhibits some properties of a excitation of the graviton
field.

\noindent {\bf iv}) The remaining contribution which absents in the flat
Lagrangian and enters the interacting Lagrangian is governed by the
contribution denoted by $\LL_8$ \rf{22112010-09}. From  \rf{22112010-09}, we
see that the tensor field $\phi_0^{ab}$ and the scalar field $\phi_0$ lead to
the appearance of a cubic potential. As is well known,  massless fields in
flat space, with exception of a scalar field, do not allow cubic vertices
without derivatives (see e.g. Ref.\cite{Metsaev:1993mj}). Also, we know that
cubic vertices having derivative independent contributions are allowed for
arbitrary spin massive fields in flat space (see e.g. \cite{Metsaev:2005ar}).
Appearance of the derivative independent cubic potential for the tensor field
$\phi_0^{ab}$ implies that the tensor field $\phi_0^{ab}$ exhibits some
features of the massive field. Taking above-given discussion into account, we
see that, on the one hand, the tensor field $\phi_0^{ab}$ exhibits some
features of excitation of the graviton field, i.e. massless spin-2 field,
and, on the other hand, the field $\phi_0^{ab}$ exhibits some features of
massive field.

{\bf Gauge transformations}. We now discuss gauge symmetries of Lagrangian
\rf{2m04122010-01}. Because we promote all gauge symmetries of free fields of
$6d$ conformal gravity  to the interacting fields we introduce the same
amount of gauge transformation parameters as in the free theory (see
\rf{2m08122010-08}),
\beq \label{2m01122010-02}
& \xi_\smminthree^a\qquad \xi_\smminone^a\qquad \xi_1^a &
\nonumber\\[-6pt]
&&
\\[-6pt]
& \xi_\smmintwo\qquad \xi_\smzero&
\nonumber
\eeq
We note that Weyl dimensions of gauge transformation parameters
\rf{2m01122010-02} are given by
\beq \label{2m01122010-08}
&& \Delta_{\xi_{k'}^a}^w= 2 + k' \,, \qquad k ' = -3,-1,1\,,
\nonumber\\[-7pt]
&&
\\[-7pt]
&& \Delta_{\xi_{k'}}^w = 2 + k'\,, \qquad k' = -2,0\,.
\nonumber
\eeq
Note also that because we are using frame-like description we assume, as
usually, the standard Lorentz gauge transformations of our fields in
\rf{2m04122010-05},
\beq
&& \delta_\lambda e_\mu^a = \lambda^{ab} e_\mu^b\,, \qquad \delta_\lambda \phi_{k'}^{ab} =
\lambda^{ac} \phi_{k'}^{cb} + \lambda^{bc} \phi_{k'}^{ac}\,,\qquad \quad k' =
0,2\,,
\nonumber\\[-7pt]
\label{2m29122010-01} &&
\\[-7pt]
&& \delta_\lambda b^a = \lambda^{ac} b^c\,,\qquad \delta_\lambda\phi_1^a =
\lambda^{ac}\phi_1^c\,,\hspace{2cm}  \delta_\lambda \phi_0 = 0 \,,
\nonumber
\eeq
$\lambda^{ab}= -\lambda^{ba}$. We now discuss gauge transformations
associated with gauge transformations parameters \rf{2m01122010-02} in turn.

{\bf $\xi_\smminthree^a$ transformation (diffeomorphism transformations)}. In
our approach, the gauge transformation parameter which is responsible for
diffeomorphism transformations is denoted by $\xi_\smminthree^a$. The
diffeomorphism transformations take the standard form
\beq
 \label{2m01122010-10}
 && \delta_{\xi_\smminthree^{\vphantom{}}}  e_\mu^a = \xi_\smminthree\partial e_\mu^a +
 e_\nu^a\partial_\mu
\xi_\smminthree^\nu\,,
\\[5pt]
\label{2m01122010-11} && \delta_{\xi_\smminthree^{\vphantom{}}}
\phi_{k'}^{ab} = \xi_\smminthree\partial \phi_{k'}^{ab}\,,\qquad k'=0,2\,,
\\[5pt]
\label{2m01122010-12}&& \delta_{\xi_\smminthree^{\vphantom{}}} b^a =
\xi_\smminthree\partial b^a\,,
\\[5pt]
\label{2m01122010-13} && \delta_{\xi_\smminthree^{\vphantom{}}} \phi_1^a =
\xi_\smminthree\partial \phi_1^a\,,
\\[5pt]
\label{2m01122010-14}&& \delta_{\xi_\smminthree^{\vphantom{}}} \phi_0 =
\xi_\smminthree\partial \phi_0\,,
\eeq
\be \xi_\smminthree\partial \equiv \xi_\smminthree^\mu \partial_\mu\,, \qquad
\xi_\smminthree^\mu \equiv e^{\mu a}\xi_\smminthree^a\,. \ee

{\bf $\xi_\smminone^a$ gauge transformations}. In our approach, gauge
transformations associated with the parameter $\xi_\smminone^a$ are
simultaneously realized as the gradient gauge transformation of the spin-2
field $\phi_0^{ab}$ and the Stueckelberg gauge transformation of the
compensator $b^a$. The $\xi_\smminone^a$ gauge transformations receive
interaction dependent corrections in the interacting theory. This is to say
that the $\xi_\smminone^a$ gauge transformations take the form
\beq
\label{2m01122010-15} && \delta_{\xi_\smminone^{\vphantom{}}} e_\mu^a = 0\,,
\\[5pt]
\label{2m01122010-16} && \delta_{\xi_\smminone^{\vphantom{}}} \phi_0^{ab} = \DD^a
\xi_\smminone^b + \DD^b \xi_\smminone^a \,,
\\[5pt]
\label{2m01122010-17} && \delta_{\xi_\smminone^{\vphantom{}}} \phi_2^{ab} =
\LL_{\xi_\smminone}\phi_0^{ab} + \phi_1^a \xi_\smminone^b + \phi_1^b
\xi_\smminone^a  - \half \eta^{ab}\phi_1^e \xi_\smminone^e\,,
\\[5pt]
\label{2m01122010-18} && \delta_{\xi_\smminone^{\vphantom{}}} b^a = -\xi_\smminone^a\,,
\\[5pt]
\label{2m01122010-19} && \delta_{\xi_\smminone^{\vphantom{}}} \phi_1^a = -\half
\phi_0^{ab}\xi_\smminone^b - \half F^{ab}\xi_\smminone^b +  \frac{u}{2}\phi_0
\xi_\smminone^a\,,
\\[5pt]
\label{2m01122010-20} && \delta_{\xi_\smminone^{\vphantom{}}} \phi_0  = 0\,,
\eeq
where the action of Lie derivative $\LL_{\xi_\smminone^{\vphantom{}}}$ on the
field $\phi_0^{ab}$ is defined to be
\be \label{2m01122010-21} \LL_{\xi_\smminone^{\vphantom{}}}\phi_0^{ab} \equiv
\xi_\smminone^c \DD^c\phi_0^{ab} + \DD^a\xi_\smminone^c \phi_0^{cb} +
\DD^b\xi_\smminone^c \phi_0^{ca}\,.\ee
Comparing these gauge transformations with free theory gauge transformations
given in \rf{2m01122010-06}, \rf{2m01122010-07},\rf{2man29112010-02}, we see
that there are two types of interaction dependent contributions. The first
ones given in \rf{2m01122010-16} are obtained simply by the covariantization
of the flat derivatives in free theory transformations, $\partial^a
\rightarrow \DD^a$, (see \rf{2m01122010-06}). The remaining contributions
given in \rf{2m01122010-17}, \rf{2m01122010-19} are obtained in due course of
building both the interacting gauge invariant Lagrangian and the
corresponding gauge transformations.

{\bf $\xi_1^a$ gauge transformations}. In our approach, gauge transformations
associated with the parameter $\xi_1^a$ are simultaneously realized as the
gradient  gauge transformation of the spin-2 field $\phi_2^{ab}$ and the
Stueckelberg gauge transformation of the vector field $\phi_1^a$. In the
interacting theory, the $\xi_1^a$ gauge transformations take the form
\beq \label{2m01122010-22}
&& \delta_{\xi_1^{\vphantom{}}} e_\mu^a = 0 \,,
\\[5pt]
\label{2m01122010-23} && \delta_{\xi_1^{\vphantom{}}} \phi_0^{ab} = 0 \,,
\\[5pt]
\label{2m01122010-24} && \delta_{\xi_1^{\vphantom{}}} \phi_2^{ab} = \DD^a \xi_1^b + \DD^b
\xi_1^a\,,
\\[5pt]
\label{2m01122010-25} && \delta_{\xi_1^{\vphantom{}}} b^a  = 0\,,
\\[5pt]
\label{2m01122010-26} && \delta_{\xi_1^{\vphantom{}}} \phi_1^a = - \xi_1^a\,,
\\[5pt]
\label{2m01122010-27} && \delta_{\xi_1^{\vphantom{}}} \phi_0 = 0 \,.
\eeq
Comparing these gauge transformations with free theory gauge transformations
given in \rf{2m01122010-07}, \rf{2man29112010-02}, we see that all that is
required for the generalization of free theory $\xi_1^a$ gauge
transformations is to make covariantization of the gradient gauge
transformation of the spin-2 field $\phi_2^{ab}$ (see \rf{2m01122010-07} and
\rf{2m01122010-24})

{\bf $\xi_\smmintwo$ gauge transformations (Weyl gauge transformations)}.  In
our approach, gauge transformation parameter responsible for Weyl gauge
transformations is denoted by $\xi_\smmintwo$. To make contact with the
commonly used notation we introduce the parameter $\sigma$ by the relation
\be \label{2m01122010-28} \sigma = - \frac{1}{4}\xi_{-2} \,.\ee
Using Weyl dimensions of our fields \rf{2m07122010-02}, we write down the
standard Weyl gauge transformations for our fields
\beq
\label{2m01122010-29} && \delta_{\xi_\smmintwo^{\vphantom{}}} e_\mu^a =   - \sigma e_\mu^a\,,
\\[5pt]
\label{2m01122010-30} && \delta_{\xi_\smmintwo^{\vphantom{}}} \phi_{k'}^{ab} =   (2+k')\sigma
\phi_{k'}^{ab}\,,\qquad k'= 0, 2\,,
\\[5pt]
\label{2m01122010-31} && \delta_{\xi_\smmintwo^{\vphantom{}}} b^a  = D^a\xi_{-2} +  \sigma b^a \,,
\\[5pt]
\label{2m01122010-32} && \delta_{\xi_\smmintwo^{\vphantom{}}} \phi_1^a =   3 \sigma \phi_1^a\,.
\\[5pt]
\label{2m01122010-33} && \delta_{\xi_\smmintwo^{\vphantom{}}} \phi_0 =   2\sigma \phi_0\,.
\eeq

{\bf $\xi_0$ gauge transformations}. In our approach, gauge transformations
associated with parameter $\xi_0$ are simultaneously realized as the gradient gauge
transformation for the spin-1 field $\phi_1^a$ and the Stueckelberg
transformation of the scalar field $\phi_0$. In the interacting theory, the
$\xi_0$ gauge transformations take the form
\beq \label{2m01122010-34}
&& \delta_{\xi_\smzero^{\vphantom{}}} e_\mu^a  = 0\,,
\\[5pt]
\label{2m01122010-35} && \delta_{\xi_\smzero^{\vphantom{}}}  \phi_0^{ab} =
\half \eta^{ab}\xi_0\,,
\\[5pt]
\label{2m01122010-36} && \delta_{\xi_\smzero^{\vphantom{}}}  \phi_2^{ab} = -
\half \phi_0^{ab}\xi_0\,,
\\[5pt]
\label{2m01122010-37} && \delta_{\xi_\smzero^{\vphantom{}}}  b^a  = 0 \,,
\\[5pt]
\label{2m01122010-38} && \delta_{\xi_\smzero^{\vphantom{}}}  \phi_1^a = \DD^a \xi_0\,,
\\[5pt]
\label{2m01122010-39} && \delta_{\xi_\smzero^{\vphantom{}}}  \phi_0 = - u \xi_0\,.
\eeq
Comparing these gauge transformations with free theory gauge transformations
given in \rf{2m01122010-07}, \rf{2man29112010-02}, we see that the
generalization of free theory $\xi_0$ gauge transformations is arrived in two
steps: i) by covariantization of the gradient gauge transformations of spin-1
field $\phi_1^a$ (see \rf{2man29112010-02} and \rf{2m01122010-38}); ii) by
modification of the gauge transformation of field $\phi_2^{ab}$ (see
\rf{2m01122010-07} and \rf{2m01122010-36}).

\bigskip
\noindent {\bf Matching of linearized background gauge symmetries of
interacting theory and global symmetries of free theory}. Let us recall the
definition of linearized background gauge symmetries. Consider a gauge
transformation for interacting field $\Phi$,
\be \delta \Phi = G_\alpha (\Phi)\xi^\alpha \,.\ee
If $\bar\Phi$ is a solution to equations of motion, then gauge transformation
that respects this solution is realized by using the gauge transformation
parameters $\bar\xi^\alpha$ satisfying the equations
\be G_\alpha(\bar\Phi)\bar\xi^\alpha = 0 \,. \ee
Using the field expansion $\Phi = \bar\Phi + \phi$, the linearized background
gauge transformations are then defined as
\be \label{2m22122010-18} \delta \phi = \partial_\Phi
G_\alpha(\bar\Phi)\bar\xi^\alpha \phi \,,\ee
where $\partial_\Phi$ stands for a functional derivative. As is well known,
the linearized background gauge transformations  are interrelated with global
symmetries of the corresponding flat theory. We now demonstrate this
interrelation for the case of $6d$ conformal gravity.

We note that solution to $6d$ conformal gravity equations of motions
corresponding to the flat space background is given by
\be \label{2m22122010-01} \bar{e}_\mu^a = \delta_\mu^a\,, \qquad \bar\phi_0^{ab} = 0\,, \qquad
\bar\phi_2^{ab} = 0 \,,\qquad \bar{b}^a =0 \,, \qquad \bar\phi_0=0\,.\ee
Collecting all gauge transformations
\be \label{2m22122010-02}
\deltabf = \delta_{\xi_\smminthree^a} + \delta_{\xi_\smminone^a} +
\delta_{\xi_1^a} +  \delta_{\xi_\smmintwo} + \delta_{\xi_0} +
\delta_{\lambda^{ab}} \ee
and using notation $\bar\Phi$ for background fields in  \rf{2m22122010-01},
we now look for gauge transformations that respect solution given in
\rf{2m22122010-01}
\be  \label{2m22122010-03}
\deltabf \bar\Phi =0\,,  \ee
To discuss solution to equations in \rf{2m22122010-03} we use the following
notation. Solution to gauge transformation parameter $\xi$ that corresponds
to symmetry generator $G$ will be denoted as $\xi^G$. In our case there are
the following set of symmetry generators $G= P^a,J^{ab},D,K^a$. We now write
solutions to gauge transformation parameters corresponding to these symmetry
generators.

\noindent {\bf Poincar\'e translations},
\beq \label{2m22122010-04}
&& \bar\xi_\smminthree^{b\,P^a}  = \eta^{ab}\,,\hspace{1.3cm}
\bar\xi_\smminone ^{b\,P^a} = 0\,, \hspace{1.3cm} \bar\xi_1^{b\,P^a} = 0\,,
\nonumber\\[-7pt]
&&
\\[-7pt]
&& \bar\xi_\smmintwo^{P^a} = 0\,, \hspace{1.8cm} \bar\xi_0^{P^a} = 0 \,,
\hspace{1.6cm} \bar\lambda^{bc\,P^a} = 0 \,;
\nonumber
\eeq
\noindent {\bf Lorentz rotations},
\beq \label{2m22122010-05}
&& \bar\xi_\smminthree^{c\,J^{ab}}  = 2\eta^{c[b} x^{a]}\,, \hspace{1.5cm}
\bar\xi_\smminone ^{c\,J^{ab}} = 0\,, \hspace{1.5cm} \bar\xi_1^{c\,J^{ab}} =
0\,,
\nonumber\\[-7pt]
&&
\\[-7pt]
&& \bar\xi_\smmintwo^{J^{ab}} = 0\,,\hspace{2.8cm}   \bar\xi_0^{J^{ab}} = 0
\,, \hspace{1.6cm} \bar\lambda^{ce,J^{ab}} = 2\eta^{a[c} \eta^{e]b}\,;
\nonumber
\eeq
\noindent {\bf Dilatation},
\beq \label{2m22122010-06}
&& \bar\xi_\smminthree^{a\,D}  = x^a\,, \hspace{1.5cm} \bar\xi_\smminone
^{a\,D} = 0\,, \hspace{1.5cm} \bar\xi_1^{a\,D} = 0\,,
\nonumber\\[-7pt]
&&
\\[-7pt]
&& \bar\xi_\smmintwo^{D} = -4\,, \hspace{1.6cm} \bar\xi_0^{D} =
0\,,\hspace{1.8cm}  \bar\lambda^{ab\,D} = 0\,;
\nonumber
\eeq
{\bf Conformal boosts},
\beq \label{2m22122010-07}
&& \bar\xi_\smminthree^{b\,K^a}  = -\half x^2 \eta^{ab} + x^a x^b\,,
\hspace{1.3cm}
\bar\xi_\smminone^{b\,K^a} = - 4 \eta^{ab}\,,
\hspace{1.3cm}
\bar\xi_1^{b\,K^a} = 0 \,,
\nonumber\\[-7pt]
&&
\\[-7pt]
&& \bar\xi_\smmintwo^{K^a} = - 4x^a\,,
\hspace{3.3cm}
\bar\xi_0^{K^a} = 0 \,,
\hspace{2.5cm}
\bar\lambda^{bc,K^a} = 2\eta^{a[b} x^{c]} \,.\qquad
\nonumber
\eeq
We now note that the linearized background gauge symmetries with gauge
transformation parameters given in \rf{2m22122010-04}, \rf{2m22122010-05},
\rf{2m22122010-06}, and \rf{2m22122010-07} correspond to the respective
Poincar\'e translation, Lorentz, dilatation, and conformal boost symmetries
of $6d$ conformal gravity in the flat space. To demonstrate this, we
introduce fields of $6d$ conformal gravity in flat space background
\rf{2m22122010-03},
\be \label{2m22122010-08} \phi_\smmintwo^{ab}\,,\quad \phi_0^{ab}\,,\quad
\phi_2^{ab}\,, \qquad  \phi_\smminone^a\,,\quad \phi_1^a\,, \qquad \phi_0 \,,
\ee
where $\phi_\smmintwo^{ab}$ appears in the small field expansion of the
vielbein field $e_\mu^a$, while $\phi_\smminone^a$ is identified with the
compensator field in flat space background,
\be \label{2m22122010-09} e_\mu^a = \delta_\mu^a + \half
\phi_\smmintwo^{ab}\eta_{\mu b}\,, \qquad\quad  b^a = \phi_\smminone^a \,.
\ee
Note that expansion for $e_\mu^a$ \rf{2m22122010-09} implies that we use
Lorentz gauge transformation \rf{2m29122010-01} to get symmetric tensor field
$\phi_\smmintwo^{ab}$. Now using gauge transformation parameters given in
\rf{2m22122010-04}, \rf{2m22122010-05}, \rf{2m22122010-06} and general
relation given in \rf{2m22122010-18} we make sure that linearized background
gauge symmetries with gauge transformation parameters in \rf{2m22122010-04},
\rf{2m22122010-05} and \rf{2m22122010-06} coincide precisely with the
respective Poincar\'e and dilatation symmetries of free $6d$ conformal theory
discussed in Sec. \ref{sec03-02}. Matching of the conformal boost symmetries
turn out to be more interesting. This is to say that the linearized
background gauge symmetries with gauge transformation parameters given in
\rf{2m22122010-07} take the form given in \rf{2m22122010-10} with the same
$K_{\Delta,M}^a$ transformations as in \rf{2m22122010-11} and the following
$R^a$ transformations:
\beq \label{2m22122010-12}
\delta_{R^a} \phi_\smmintwo^{bc} & = & 0 \,,
\\[5pt]
\label{2m22122010-13} \delta_{R^a} \phi_0^{bc}  & =  & - 2\eta^{ab} \phi_\smminone^c -2 \eta^{ac}
\phi_\smminone^b + 2 \eta^{bc} \phi_\smminone^a
\nonumber\\[5pt]
& - & 4 \partial^a \phi_{-2}^{bc} + 2\partial^b \phi_{-2}^{ac}  + 2\partial^c
\phi_{-2}^{ab}\,,
\\[5pt]
\label{2m22122010-14} \delta_{R^a} \phi_2^{bc}  & = &  -  4 \eta^{ab} \phi_1^c - 4\eta^{ac}\phi_1^b
+ 2\eta^{bc}\phi_1^a - 4 \partial^a \phi_0^{bc}\,,
\\[5pt]
\label{2m22122010-15} \delta_{ R^a} \phi_\smminone^b & = &  2\phi_\smmintwo^{ab}\,,
\\[5pt]
\label{2m22122010-16} \delta_{R^a} \phi_1^b & =  & 2\phi_0^{ab}  - 2\eta^{ab} u\phi_0 -
2F^{ab}(\phi_\smminone) \,,
\\[5pt]
\label{2m22122010-17} \delta_{R^a} \phi_0  & = & 0\,.
\eeq
Comparing $R^a$ transformations in \rf{24122009-01}-\rf{24122009-06} and the
ones in \rf{2m22122010-12}-\rf{2m22122010-17}, we notice some differences in
$R^a$ transformations for the fields $\phi_0^{bc}$, $\phi_{-1}^b$,
$\phi_1^b$, and $\phi_0$. Explanation of these differences is obvious: global
transformations of gauge fields are defined up to gauge transformations.
Introducing notation for the gauge transformation parameters
\be
\xi_{-1}^{aK^b} \equiv 2\phi_{-2}^{ab}\,,
\qquad \qquad \xi_0^{K^a} = 2 \phi_\smminone^a\,,
\ee
and using notation $\delta_{R^a}^\flat$ and $\delta_{R^a}$ for the respective
$R^a$ transformation in \rf{24122009-01}-\rf{24122009-06} and
\rf{2m22122010-12}-\rf{2m22122010-17}, we note that $R^a$ transformations
given in \rf{24122009-01}-\rf{24122009-06} and the ones in
\rf{2m22122010-12}-\rf{2m22122010-17} are related by the gauge
transformations,
\beq
&& \delta_{R^a} \phi_0^{bc} = \delta_{R^a}^\flat \phi_0^{bc} +
\partial^b \xi_{-1}^{cK^a}  + \partial^c \xi_{-1}^{bK^a} + \half
\eta^{bc}\xi_0^{K^a}\,,
\\[5pt]
&& \delta_{R^a}\phi_{-1}^b = \delta_{R^a}^\flat \phi_{-1}^b - \xi_{-1}^{b
K^a}\,,
\\[5pt]
&& \delta_{R^a}\phi_1^a = \delta_{R^a}^\flat \phi_1^b +
\partial^b \xi_0^{K^a}\,,
\\[5pt]
&& \delta_{R^a} \phi_0 = \delta_{R^a}^\flat\phi_0 - u \xi_0^{K^a}\,.
\eeq
Thus, we see that the conformal boost transformations also match.

\newsection{ Higher-derivative Lagrangian of interacting $6d$ conformal gravity
}\label{sec-04-a1}

Our ordinary-derivative Lagrangian can be used for the derivation of the
higher-derivative Lagrangian of interacting $6d$ conformal gravity. The
higher-derivative Lagrangian of interacting theory can be obtained by
following the procedure we used for the derivation of the higher-derivative
Lagrangian of free $6d$ conformal gravity in Sec.\rf{sec03-02}. We now to
proceed to details of the derivation.

From gauge transformations
\rf{2m01122010-18},\rf{2m01122010-26},\rf{2m01122010-39}, we see that the
vector fields $b^a$, $\phi_1^a$ and the scalar field $\phi_0$ transform as
Stueckelberg fields and can therefore be gauged away by fixing the
Stueckelberg gauge symmetries. Gauging away the vector fields and the scalar
field,
\be \label{2m01122010-40} b^a = 0 \,,\qquad \phi_1^a = 0 \,,\qquad \phi_0 = 0
\,,\ee
we see that our Lagrangian \rf{2m04122010-01} takes the simplified form
\beq
&& \hspace{-1cm} \label{2m01122010-41} \LL = \LL_1 + \LL_2 + \LL_6 + \LL_8\,,
\
\\[5pt]
\label{2m01122010-42} && e^{-1} \LL_1 = -\phi_2^{ab} G^{ab} \,,
\\[5pt]
\label{2m01122010-43} && e^{-1}\LL_2 = - \frac{1}{4}D^a \phi_0^{bc}D^a \phi_0^{bc} +
\frac{1}{8}D^a \phi_0^{bb} D^a \phi_0^{cc} + \half C_1^a C_1^a
\nonumber\\[5pt]
&& \hspace{1.3cm}  - \,\, \half R^{cabe} \phi_0^{ab} \phi_0^{ce} + \half
R^{ab} \phi_0^{ac}\phi_0^{cb} -  \half R^{ab} \phi_0^{ab} \phi_0^{cc} +
(\frac{1}{8} \phi_0^{aa}\phi_0^{bb} - \frac{1}{4} \phi_0^{ab}\phi_0^{ab})
R\,,\qquad
\\[5pt]
\label{2m01122010-44} && e^{-1} \LL_6 = -\half \phi_2^{ab}\chi_0^{ab}\,,
\\[5pt]
\label{2m01122010-45} &&  e^{-1} \LL_8 =
\frac{1}{4}\phi_0^{ab}\phi_0^{bc}\phi_0^{ca}
-\frac{5}{16}\phi_0^{ab}\phi_0^{ab}\phi_0^{cc} +
\frac{1}{16}(\phi_0^{aa})^3\,,
\\[5pt]
\label{2m01122010-46} && \hspace{2cm} G^{ab} \equiv R^{ab} - \half
\eta^{ab}R\,,
\\[5pt]
\label{2m01122010-47} && \hspace{2cm} C_1^a \equiv D^b \phi_0^{ab} - \half
D^a \phi_0^{cc}\,,
\\[5pt]
\label{2m01122010-48} && \hspace{2cm} \chi_0^{ab} \equiv \phi_0^{ab} -
\eta^{ab} \phi_0^{cc} \,.
\eeq
We note that the Lagrangian \rf{2m01122010-41} depends on the rank-2 tensor
field $\phi_2^{ab}$ linearly (see expressions for $\LL_1$ and $\LL_6$). Using
equations of motion for the field $\phi_2^{ab}$ obtained from Lagrangian
\rf{2m01122010-41} we find the equation
\be \label{2m01122010-49} \phi_0^{ab} - \eta^{ab}\phi_0^{cc} = - 2G^{ab}\,,
\ee
which has obvious solution
\be \label{2m01122010-50} \bar\phi_0^{ab} = - 2 R^{ab} + \frac{1}{5}\eta^{ab}
R\,. \ee
Plugging solution $\bar\phi_0^{ab}$ \rf{2m01122010-50} into Lagrangian
\rf{2m01122010-41} we obtain the higher-derivative Lagrangian
\be \label{2m02122010-01}
e^{-1}\LL = R^{ab}D^2 R^{ab} - \frac{3}{10}R D^2 R - 2 R^{cabe}R^{ab}R^{ce} -
R^{ab}R^{ab} R + \frac{3}{25}R^3\,.
\ee
This higher-derivative Lagrangian should be invariant under Weyl gauge
transformations
\be \label{2m02122010-02} \delta e_\mu^a = -\sigma e_\mu^a \,.\ee
We have checked directly that the Lagrangian is indeed invariant under Weyl
gauge transformations \rf{2m02122010-02}. Note that by using various
identities for $R^3$ terms (see e.g. Ref.\cite{Metsaev:1986yb}), the
Lagrangian can be expressed in terms of the  Weyl tensor and Ricci
curvatures.

Though there is a lot of  literature on  conformal gravity, we did not find
discussion of the Lagrangian \rf{2m02122010-01} in the earlier literature. We
note however that all Weyl invariant densities for $6d$ conformal theory were
presented in Ref.\cite{Bonora:1985cq} (see also
Refs.\cite{Deser:1993yx,Karakhanian:1994yd,Erdmenger:1997gy}%
\footnote{ Discussion of interesting methods for constructing Weyl invariant
densities may be found in \cite{Boulanger:2004zf,Boulanger:2004eh}.
Classification of all the six-derivative Lagrangians in arbitrary dimensions
such that the trace of the resulting field equations are at most of order 3
may be found in Ref.\cite{Oliva:2010zd}. Study of the effective $6d$
conformal gravity may be found in Ref.\cite{Odintsov:1994vu}.}).
Up to total derivative, in $6d$ conformal gravity theory, there are three Weyl
invariant densities constructed out the Weyl tensor, Ricci curvatures,
and covariant derivative. In Ref.\cite{Bonora:1985cq}, the simplest
combination of those three invariants, which involves $R^{ab}D^2R^{ab}$ term,
has been found (see Eq.(4.12) in Ref.\cite{Bonora:1985cq}). It turns out that
it is this simplest invariant that coincides
with our Lagrangian in \rf{2m02122010-01}%
\footnote{ In \rf{2m02122010-01}, our signs in front of terms involving odd
number of the Ricci tensors and Ricci scalars are opposite to the ones in
Ref.\cite{Bonora:1985cq}. Perhaps these sign differences can be explained by
the different conventions used for the definition of the Ricci tensor and
Ricci scalar in our paper and in Ref.\cite{Bonora:1985cq}. Our curvature
conventions are $R^\lambda{}_{\mu\nu\rho} =
\partial_\nu\Gamma_{\mu\rho}^\lambda-\ldots$,
$R_{\mu\nu}=R^\lambda{}_{\mu\lambda\nu}$, $R=R_\mu^\mu$.}.

As a side of remark we note that the remaining two Weyl invariant densities
also can be lifted to our gauge-invariant approach, i.e., it is possible to
build the invariants which respect all gauge symmetries of our approach. To
this end we introduce the new curvature

\beq \label{2m08122010-09}
&& \RR^{abce} = R^{abce} + \eta^{ac} \psi^{be} -  \eta^{bc} \psi^{ae}  +
\eta^{be} \psi^{ac} - \eta^{ae} \psi^{bc} \,,
\\[5pt]
\label{2m08122010-10} && \hspace{1cm} \psi^{ab}  \equiv \frac{q}{2}\Bigl(
\phi_0^{ab} + D^a b^b + D^b b^a + 2q b^a b^b - q \eta^{ab} b^2\Bigr) \,,
\eeq
$q=1/4$, and note that under $\xi_\smmintwo$, $\xi_\smminone^a$, and
$\xi_1^a$ gauge transformations new curvature $\RR^{abce}$ \rf{2m08122010-09}
transforms as
\be \label{2m08122010-11} \delta_{\xi_\smmintwo} \RR^{abce} = 2\sigma
\RR^{abce} \,, \qquad \delta_{\xi_\smminone} \RR^{abce} = 0 \,, \qquad
\delta_{\xi_1} \RR^{abce} = 0 \,, \ee

\be \label{2m08122010-12} \delta_{\xi_0} \RR^{abce}
=\frac{q}{2}(\eta^{ac}\eta^{be} - \eta^{ae}\eta^{bc})\xi_0 \,.\ee
From these relations, we see that the curvature $\RR^{abce}$ has Weyl dimension equal to 2
and this curvature is invariant under the $\xi_\smminone^a$ and $\xi_1^a$ gauge transformations. Also
we note that gauging away Stueckelberg fields \rf{2m01122010-40} and using
solution \rf{2m01122010-50}, the curvature $\RR^{abce}$ becomes the standard
Weyl tensor
\be \label{2m08122010-13}
\RR^{abce}|_{b^a=0\,,\phi_0^{ab} = \bar\phi_0^{ab}} = C^{abce} \,. \ee
We see however that the curvature $\RR^{abce}$ does not respect $\xi_0$ gauge
symmetry \rf{2m08122010-12}. General curvature that respects the $\xi_0$
gauge symmetry can be built as follows
\beq \label{2m09122010-01}
\Rbf^{abce} &=& \RR^{abce} + h_1 (\eta^{ac}\RR^{be} - \eta^{bc}\RR^{ae} +
\eta^{be}\RR^{ac} - \eta^{ae}\RR^{bc})
\nonumber\\[5pt]
&+& h_2 (\eta^{ac} \eta^{be} - \eta^{ae} \eta^{bc})\RR + h_3 (\eta^{ac}
\eta^{be} - \eta^{ae} \eta^{bc})\phi_0 \,,
\eeq
where $\RR^{ab} = \RR^{cacb}$, $\RR = \RR^{aa}$ and coefficients $h_1$,
$h_2$, $h_3$ satisfy the equation
\be \label{2m09122010-02} 1 + 10 h_1 + 30 h_2 - \frac{2u}{q}h_3 = 0 \,, \ee
and $u$ is given in \rf{2m01122010-03}. Equation \rf{2m09122010-02} is simply obtained by requiring the curvature
$\Rbf^{abce}$ to be invariant under $\xi_0$ gauge transformations
\rf{2m01122010-34}-\rf{2m01122010-39}. Thus curvature \rf{2m09122010-01} has
the desired properties
\be \label{2m08122010-16} \delta_{\xi_\smmintwo} \Rbf^{abce} = 2\sigma
\Rbf^{abce} \,, \qquad \delta_{\xi_\smminone} \Rbf^{abce} = 0 \,, \qquad
\delta_{\xi_1} \Rbf^{abce} = 0 \,, \qquad \delta_{\xi_0} \Rbf^{abce} = 0 \,,
\ee
Using the curvature $\Rbf^{abce}$, we can construct the remaining two gauge invariant
densities in a straightforward way. For instance, we can consider the
invariants
\be \label{2m08122010-17} e\Rbf_{ce}^{ab}\Rbf_{fg}^{ce}\Rbf_{ab}^{fg}\,,
\quad\qquad e\Rbf_{ef}^{ab} \Rbf_{fg}^{bc} \Rbf_{ge}^{ca}\,.\ee
In our gauge invariant approach, these invariants are counterparts of the
well known Weyl invariants appearing in the standard approach to $6d$
conformal theory,
\be \label{2m08122010-18} eC_{ce}^{ab}C_{fg}^{ce}C_{ab}^{fg}\,, \quad\qquad
eC_{ef}^{ab} C_{fg}^{bc} C_{ge}^{ca}\,,\ee
i.e. by using Stueckelberg gauge conditions \rf{2m01122010-40}  and pugging
solution \rf{2m01122010-50} into invariants \rf{2m08122010-17}, we obtain the
respective Weyl invariant densities given in \rf{2m08122010-18}.

Thus, the remaining two invariants \rf{2m08122010-17} respect all gauge
symmetries of our gauge invariant approach. The important difference of these
two invariants as compared to our Lagrangian \rf{2m04122010-01} is that these two
invariants involve the higher-derivatives, while our Lagrangian involves only
the ordinary derivatives.%
\footnote{ In Ref.\cite{Boulanger:2001he} (see Sec.8.2), authors conjectured
that two invariants given in \rf{2m08122010-18} do not deform gauge algebra
in the standard higher-derivative approach to conformal $6d$ gravity. The
fact that our remaining two invariants \rf{2m08122010-17} respect our
Stueckelberg gauge symmetries seems to be in agreement with this conjecture.}
At present time,  we do not know representation of the invariants
\rf{2m08122010-17} in terms of the ordinary derivatives. It is not obvious
that such representation can be constructed without adding new fields to our
field content in \rf{2m04122010-05}.

We finish with remark that some special representatives of general curvature
\rf{2m09122010-01} might be interesting in various contexts. For instance,
solution to equation \rf{2m09122010-02} given by
\be h_1=-\frac{1}{4}\,, \qquad h_2 = \frac{1}{20}\,, \qquad h_3 = 0\,, \ee
leads to traceless tensor
\beq \label{2m08122010-14}
\Rbf_\weyl^{abce} &=& \RR^{abce} - \frac{1}{4}(\eta^{ac}\RR^{be} -
\eta^{bc}\RR^{ae} + \eta^{be}\RR^{ac} - \eta^{ae}\RR^{bc})
\nonumber\\
& + & \frac{1}{20}(\eta^{ac} \eta^{be} - \eta^{ae} \eta^{bc}) \RR\,,
\eeq
which can be considered as counterpart of the Weyl tensor in our approach.
Another solution to equation \rf{2m09122010-02} given by
\be \label{2m09122010-03} h_1 = 0 \,,\qquad h_2 = 0\,, \qquad h_3 =
\frac{q}{2u} \,, \ee
leads to the curvature
\be \label{2m08122010-15} \Rbf_\scal^{abce} = \RR^{abce} +
\frac{q}{2u}(\eta^{ac}\eta^{be} - \eta^{ae}\eta^{bc})\phi_0\,.
\ee
The curvature $\Rbf_\scal^{abce}$ has the following interesting property. If
we introduce the corresponding Einstein tensor
\be \label{2m09122010-04} \Gbf_\scal^{ab} = \Rbf_\scal^{ab} - \half \eta^{ab}
\Rbf_\scal\,, \ee
where $\Rbf_\scal^{ab} = \Rbf_\scal^{cacb}$, $\Rbf_\scal = \Rbf_\scal^{aa}$,
then it turns out that the $\LL_1$ and $\LL_6$ parts of the Lagrangian
\rf{2m04122010-01} can be collected as
\be \LL_1 + \LL_6 = - \phi_2^{ab} \Gbf_\scal^{(ab)} \,. \ee
Because the field $\phi_2^{ab}$ does not appear in the remaining
contributions to Lagrangian \rf{2m04122010-01}, equations of motion for this field can be
represented as
\be \Gbf_\scal^{(ab)}= 0 \,.\ee

\newsection{Conclusions}\label{conl-sec-01}

In this paper, we applied the ordinary-derivative approach, developed in
Ref.\cite{Metsaev:2007fq}, to the study of $6d$ interacting conformal
gravity. The results presented here should have a number of interesting
applications and generalizations. Let us comment on some of them.

i) As we have already mentioned,  the gauge symmetries of our Lagrangian make
it possible to match  our approach with  the standard one, i.e., by an
appropriate gauge fixing of the Stueckelberg fields and solving the
constraints, we obtain the higher-derivative formulation of the $6d$ conformal
gravity. This implies that, at least at the classical level, our $6d$
conformal gravity theory is equivalent to the standard one. In this respect
it would be interesting to investigate quantum equivalence of our theory and
the standard one. We note that our formulation provides new interesting
possibilities for investigation of quantum behavior of conformal gravity. The
first step in studying quantum behavior of conformal gravity is a computation
of one-loop effective action. One  powerful method  of  the computation  of
the one-loop effective action of Einstein  gravity is based on the use of so
called $\Delta_2$ algorithm
\cite{'tHooft:1974bx,Gilkey:1975iq,Barvinsky:1985an} (see also
\cite{Fradkin:1982kf}%
\footnote{ Generalization of methods in
Refs.\cite{'tHooft:1974bx,Gilkey:1975iq,Barvinsky:1985an} to quantum
effective actions for brane induced gravity models may be found in
Refs.\cite{Barvinsky:2005ms}.}).
In order to investigate quantum properties of fourth-derivative $4d$ Weyl
gravity this algorithm was generalized to the so called $\Delta_4$-algorithm
(see \cite{Fradkin:1981iu} and refs. there). We note, however,  that since
our approach does not involve higher-derivatives and formulated in terms of
conventional kinetic terms we can use the standard $\Delta_2$ algorithm for
the investigation of the one-loop effective action.

ii) In addition to the local Weyl and diffeomorphism symmetries that enter
the standard approach to conformal gravity our approach involves gauge
symmetries for two rank-2 tensor fields and some amount of Stueckelberg gauge
symmetries. In other words, we deal with extended gauge algebra. In this
respect,  it would be interesting to analyze the general solution of the
Wess-Zumino consistency condition for our gauge algebra along the lines in
Ref.\cite{Boulanger:2004eh}.

iii) Results in this paper provide
the complete ordinary-derivative description of interacting $6d$ conformal gravity. It would be
interesting to apply these results to the study of supersymmetric conformal
field theories \cite{Bergshoeff:1980is}-\cite{Bergshoeff:1986wc} in the
framework of ordinary-derivative approach. The first step in this direction
would be understanding of how the supersymmetries are realized in the
framework of our approach.

iv) BRST approach is one of powerful approaches to the analysis of various
aspects of relativistic dynamics (see e.g.
Refs.\cite{Siegel:1999ew}-\cite{Polyakov:2009pk}). This approach turned out
to be successful for application to string theory. We believe therefore that
use of this approach for the study of conformal fields might also be helpful
for the better understanding of conformal gravity theory.

v) In the last years, there were interesting developments in the studying
mixed-symmetry fields \cite{Metsaev:1995re}-\cite{Alkalaev:2010af} that are
invariant with respect to anti-de Sitter or Minkowski space-time symmetries.
It would be interesting to apply methods developed in
Refs.\cite{Metsaev:1995re}-\cite{Alkalaev:2010af} to the studying
interacting $6d$ conformal mixed-symmetry fields%
\footnote{ Unfolded form of equations of motion for conformal mixed-symmetry
fields is studied in Ref.\cite{Shaynkman:2004vu}. Higher-derivative
Lagrangian formulation of the mixed-symmetry conformal fields was recently
developed in Ref.\cite{Vasiliev:2009ck}.}.
There are other various interesting approaches in the literature which could
be used to discuss the ordinary-derivative formulation of $6d$ conformal
fields. This is to say that various recently developed interesting
formulations in terms of unconstrained fields in flat space may be found e.g.
in Refs.\cite{Buchbinder:2008ss}-\cite{Campoleoni:2008jq}.

\bigskip

{\bf Acknowledgments}. This work was supported by the RFBR Grant
No.08-02-00963, by the Dynasty Foundation and by the Alexander von Humboldt
Foundation Grant PHYS0167.

\setcounter{section}{0}\setcounter{subsection}{0}
\appendix{ Notation }

{\bf Flat space-time notation}. Our conventions are as follows. Coordinates
in the flat space-time are denoted by $x^a$, while $\partial_a$ stands for
derivative with respect to $x^a$, $\partial_a \equiv \partial / \partial
x^a$. Vector indices of the Lorentz algebra $so(5,1)$ take the values
$a,b,c,e=0,1,\ldots ,5$. To simplify our expressions we drop the flat metric
$\eta_{ab}$ in scalar products, i.e. we use $X^aY^a \equiv \eta_{ab}X^a Y^b$.
We use operators constructed out of the coordinates and derivatives,
\be \Box=\partial^a\partial^a\,,\qquad  x\partial =x^a\partial^a\,.\ee

\noindent {\bf Curved space-time notation}. We use space-time base manifold
indices $\mu,\nu,\rho,\sigma = 0,1,\ldots ,5$ and tangent-flat vectors
indices of the $so(5,1)$ algebra $a,b,c,e,f = 0,1,\ldots ,5$. Base manifold
coordinates are denoted by $x^\mu$, while $\partial_\mu$ denote the
respective derivatives, $\partial_\mu \equiv \partial / \partial x^\mu$. We
use notation $e_\mu^a$ and $\omega_\mu^{ab}(e)$ for the respective vielbein
and the Lorentz connection. Contravariant tensor field carrying base manifold
indices, $\phi^{\mu_1\ldots \mu_s}$, is related to tensor field carrying the
tangent-flat indices, $\phi^{a_1\ldots a_s}$, in a standard way
$\phi^{a_1\ldots a_s} \equiv e_{\mu_1}^{a_1}\ldots e_{\mu_s}^{a_s}
\phi^{\mu_1\ldots \mu_s}$. Covariant derivative $D_\mu$, acting on vector
field
\be D_\mu \phi^a = \partial_\mu \phi^a + \omega_\mu^{ab}(e)\phi^b\,, \ee
satisfies the standard commutator
\be [D_\mu,D_\nu] \phi^a = R_{\mu\nu}{}^{ab}\phi^b \,.\ee
Instead of $D_\mu$, we
prefer to use a covariant derivative with the flat indices $D^a$,
\be
D_a \equiv e_a^\mu D_\mu\,,\qquad D^a = \eta^{ab}D_b\,,\ee
where $e_a^\mu$ is inverse of the vielbein, $e_\mu^a e_b^\mu = \delta_b^a$.

For field $\phi^a$ with Weyl dimension $\Delta_\phi^w$, covariant derivative
defined in \rf{2m08122010-01} satisfies the commutator
\be [\DD^a,\DD^b]\phi^c = \Rwh^{abce}\phi^e + q \Delta_\phi^w F^{ab}(b)\phi^c
\,,\qquad  \quad q=1/4\,,   \ee
where the shifted curvature and field strength $F^{ab}$ are defined in
\rf{2m04122010-04a1} and \rf{22112010-11a1} respectively. Weyl gauge
transformations defined in \rf{2m01122010-28}-\rf{2m01122010-33} lead to the
following transformation rules:
\beq
&& \delta \omega^{abc} = \sigma \omega^{abc} + \eta^{ac} D^b\sigma -
\eta^{ab} D^c\sigma\,,
\\[5pt]
&& \delta \Rwh^{abce} = 2\sigma \Rwh^{abce}  \,,
\\[5pt]
&& \delta \omegawh^{abc} = \sigma \omegawh^{abc}\,,
\\[5pt]
&& \delta \FF^{ab}(\phi) = (\Delta_{\phi}^w+1) \sigma \FF^{ab}(\phi)\,,
\\[5pt]
&& \delta F^{ab} = 2 \sigma F^{ab}\,,
\\[5pt]
&& \delta (\DD^a \phi^{a_1\ldots a_s}) = (\Delta_\phi^w+1)
\DD^a\phi^{a_1\ldots a_s}\,.
\eeq
These relations imply the following Weyl dimensions:
\be \Delta_{\Rwh^{abce}}^w = 2\,, \qquad \Delta_{\omegawh^{abc}}^w = 1
\,,\qquad \Delta_{F^{ab}}^w = 2\,,\qquad \Delta_{\FF^{ab}(\phi)}^w =
\Delta_\phi^w+1\,. \ee

Bianchi identities for the shifted curvature and field strength
\rf{22112010-11a1},
\beq
&& \label{29122009-02} \DD^f \Rwh^{abce} + \hbox{ cycl.perms.} (fab) = 0
\,,
\\[5pt]
&& \label{29122009-03} \DD^a F^{bc} + \hbox{ cycl.perms.} (abc) = 0 \,,\eeq
can be obtained in a usual way. Using explicit relations for the curvatures
and the Einstein tensor
\beq \label{2m17122010-09}
&& \Rwh^{abce} = R^{abce} + \eta^{ac}\varphi^{be} - \eta^{bc}\varphi^{ae}  +
\eta^{be}\varphi^{ac} - \eta^{ae}\varphi^{bc}\,,
\\[5pt]
\label{2m17122010-10} && \Rwh^{ab} = \Rwh^{cacb}\,, \qquad  \Rwh =
\Rwh^{aa}\,,
\\[5pt]
\label{2m17122010-11} && \Rwh^{ab} = R^{ab} + 4 \varphi^{ab} +
\eta^{ab}\varphi^{cc} \,,
\\[5pt]
\label{2m17122010-12} && \Rwh = R + 10\varphi^{aa} \,,
\\[5pt]
&& \Gwh^{ab} \equiv \Rwh^{ab} - \half \eta^{ab}\Rwh\,,
\\[5pt]
\label{2m17122010-13} && \varphi^{ab} = q D^a b^b + q^2 b^a b^b - \half q^2
\eta^{ab} b^2\,,
\\[5pt]
\label{2m17122010-14} && \varphi^{cc} = q D b - 2 q^2  b^2\,, \qquad q=1/4\,,
\eeq
we obtain various useful identities
\beq
&& \Rwh^{abce} - \Rwh^{ceab} = q (\eta^{ac} F^{be} -  \eta^{bc} F^{ae}  +
\eta^{be} F^{ac} -  \eta^{ae} F^{bc}) \,,
\\[5pt]
&& \Rwh^{ab} - \Rwh^{ba} =  F^{ab}\,,
\\[5pt]
&& \DD^c \Rwh^{abce}  =  \DD^a \Rwh^{be} - \DD^b \Rwh^{ae} \,,
\\[5pt]
&& \DD^c \Rwh^{ceab}  = \DD^a \Rwh^{be} - \DD^b \Rwh^{ae}
\nonumber\\[5pt]
&& \hspace{2cm} + \,\, q(\DD^e F^{ab} -  \eta^{ae} \DD^c F^{cb} + \eta^{be} \DD^c F^{ca})\,,
\\[5pt]
\label{2m08122010-03}
&& \DD^b \Rwh^{ab} = \half \DD^a \Rwh\,,
\\[5pt]
\label{2m08122010-04} && \DD^b \Rwh^{ba} = \half \DD^a \Rwh + \DD^b F^{ba}\,,
\\[5pt]
\label{2m08122010-05}
&& \DD^b \Gwh^{ab} = 0\,,
\qquad
\DD^b \Gwh^{ba} = \DD^b F^{ba}\,,
\qquad
\DD^b \Gwh^{(ab)} = \half \DD^b F^{ba}\,.
\eeq
Throughout this paper, symmetrization and antisymmetrization of the indices
are normalized as $(ab) = \half ( ab + ba)$, $[ab] = \half ( ab - ba)$.

\appendix{ Derivation of interacting gauge invariant Lagrangian }

In this Appendix we outline some details of the derivation of gauge invariant
Lagrangian given in \rf{2m04122010-01} and the corresponding gauge
transformations. We divide our derivation in seven steps which we now discuss
in turn.

{\bf Step 1}. We begin with the discussion of contribution to Lagrangian
\rf{2m04122010-01}  denoted by $\LL_1$ \rf{22112010-01}. This contribution is
simply obtained from the contribution to free theory theory Lagrangian
\rf{2m04122010-01a1} also denoted by $\LL_1$ \rf{2m04122010-01a2} in a
straightforward way. Namely $\LL_1$ \rf{22112010-01} is obtained from $\LL_1$
\rf{2m04122010-01a2} by requiring $\LL_1$ \rf{22112010-01} to be invariant
under Weyl gauge transformations given in
\rf{2m01122010-29}-\rf{2m01122010-33}. All that is required to respect those
gauge transformations is to replace the linearized shifted Einstein tensor
and curvatures given in \rf{2m17122010-02}-\rf{2m17122010-07} by the
corresponding complete shifted Einstein tensor and curvatures given in
\rf{2m17122010-08} and \rf{2m17122010-11},\rf{2m17122010-12}.

{\bf Step 2}. We covariantize the flat $\xi_\smminone^a$ gauge transformation
of the field $\phi_0^{ab}$ \rf{2m01122010-06} by making replacement
$\partial^a \rightarrow \DD^a$, while the $\xi_\smminone^a$ gauge
transformation of the field $b^a$ \rf{2man29112010-01} is not changed,
\beq \label{22112010-16a1}
&&  \delta_{\xi_{-1}} \phi_0^{ab} = \DD^a \xi_\smminone^b + \DD^b
\xi_\smminone^a \,,
\\[5pt]
\label{22112010-16a2}
&& \delta_{\xi_{-1}}b^a = -\xi_{-1}^a \,.
\eeq
Also, making the covariantization $\partial^a \rightarrow \DD^a$ in
contribution to flat Lagrangian denoted by $\LL_2$ \rf{2m04122010-01a3}, we
introduce
\be
\label{2m18122010-01} e^{-1}\LL_{2\smK} = - \frac{1}{4}\DD^a \phi_0^{bc}\DD^a
\phi_0^{bc} + \frac{1}{8}\DD^a \phi_0^{bb}\DD^a \phi_0^{cc} + \half C_1^a
C_1^a\,,
\ee
where the covariantized $C_1^a$ is defined as in \rf{22112010-14}. After
this, we consider gauge variation of $\LL_{2\smK}$ under gauge
transformations \rf{22112010-16a1},\rf{22112010-16a2}. In the gauge
variation, we find unwanted terms proportional to the shifted curvatures
$\Rwh^{abce}$, $\Rwh^{ab}$, $\Rwh$ which cannot be cancelled by modification
of gauge transformations of the field entering our field content
\rf{2m04122010-05}. Our observation is that those unwanted terms can be
cancelled by adding to $\LL_{2\smK}$ the following contribution
\be \label{2m18122010-02}
e^{-1}\LL_{2\smR} = - \half \Rwh^{cabe} \phi_0^{ab} \phi_0^{ce} + \half
\Rwh^{ab} \phi_0^{ac}\phi_0^{cb} -  \half \Rwh^{ab} \phi_0^{ab} \phi_0^{cc} +
(\frac{1}{8} \phi_0^{aa}\phi_0^{bb} - \frac{1}{4} \phi_0^{ab}\phi_0^{ab})
\Rwh\,.
\ee
This is to say that variation of $\LL_2 = \LL_{\smK} + \LL_{2\smR}$ under
gauge transformations \rf{22112010-16a1},\rf{22112010-16a2} takes the form
\beq \label{22112010-16a3}
&& \delta_{\xi_{-1}}\LL_2 = \delta_{{\phi_0^{ab}},\xi_{-1}} \LL_2  +
\delta_{b^a, \xi_{-1}}\LL_2\,,
\\[7pt]
\label{22112010-16a4} &&\hspace{1.7cm}  \delta_{\phi_0^{ab}, \xi_{-1}} \LL_2  =
\delta_{\phi_0^{ab}, \xi_{-1}}\LL_2\bigr|_{\Gwh} + \delta_{\phi_0^{ab},
\xi_{-1}} \LL_2 \bigr|_F \,, \eeq

\beq \label{22112010-16a5}
e^{-1}\delta_{\phi_0^{ab},\xi_{-1}} \LL_2\bigr|_{\Gwh} & = & \Gwh^{ab}
 \LL_{\xi_\smminone^{\vphantom{}}}\phi_0^{ab}\,,
\\[5pt]
\label{22112010-16a6} e^{-1} \delta_{\phi_0^{ab},\xi_{-1}} \LL_2\bigr|_F & = & - 2q
F^{ac}\FF^{cb}(\xi_\smminone) \phi_0^{ab} + 2q F^{ab}\xi_\smminone^b
(\DD\phi_0)^a - q F^{ab} \xi_\smminone^b \DD^a\phi_0^{cc}
\nonumber\\[5pt]
& + & 2q \DD^a F^{ab}\xi_\smminone^c \phi_0^{bc} + q \DD^a
F^{ab}\xi_\smminone^b \phi_0^{cc}\,,
\\[5pt]
\label{22112010-16a7}
e^{-1}\delta_{b^a,\xi_{-1}} \LL_2  & = & 2q \xi^b\phi_0^{ab} ( \DD^c\phi_0^{ac} -
\DD^a\phi_0^{cc})
\nonumber\\[5pt]
& - & 2q \DD^a \xi^b (\phi^2)^{ab} -
\frac{q}{2}\phi_0^{aa}\phi_0^{bb}\DD\xi + \frac{3q}{2}
\phi_0^{ab}\phi_0^{ab}\DD\xi + q\DD^a \xi^b \phi_0^{ab}\phi_0^{cc}\,,\qquad
\eeq
where Lie derivative $\LL_{\xi_\smminone^{\vphantom{}}}\phi_0^{ab}$ entering
variation \rf{22112010-16a5} is defined in \rf{2m01122010-21}. From these
relations, we see that the remaining terms involving the shifted curvatures
are proportional to $\Gwh^{ab}$ \rf{22112010-16a5}. Such terms can easily be
cancelled by modifying gauge transformation of the field $\phi_2^{ab}$,
\be \label{2m18122010-03} \delta_{\xi_\smminone}'\phi_2^{ab}
= \LL_{\xi_\smminone^{\vphantom{}}}\phi_0^{ab}\,. \ee
Namely, it is easy to see that variation of $\LL_1$ \rf{22112010-01} under
gauge transformation of $\phi_2^{ab}$ given in \rf{2m18122010-03} cancels
variation in \rf{22112010-16a5}.

{\bf Step 3}. We now consider variation of contributions to the Lagrangian
denoted by $\LL_3$ \rf{22112010-03} and $\LL_5$ \rf{22112010-05} under gauge
transformations given in \rf{22112010-16a1},\rf{22112010-16a2}. Using
notation $\delta_{b^a,\xi_{-1}}\LL_3$ and $\delta_{b^a,\xi_{-1}}\LL_5$ for
the respective variations of $\LL_3$ and $\LL_5$ under $\xi_\smminone^a$
gauge transformation of the field $b^a$, and notation
$\delta_{\phi_0^{ab},\xi_{-1}}\LL_5 $ for variation of $\LL_5$ under
$\xi_\smminone^a$ gauge transformation of the field $\phi_0^{ab}$ we find
\beq \label{2m18122010-06}
e^{-1}(\delta_{b^a,\xi_{-1}}\LL_3 + \delta_{\phi_0^{ab},\xi_{-1}}\LL_5) & = &
(\phi_1^a \xi_{-1}^b + \phi_1^b \xi_{-1}^a - \frac{1}{2} \eta^{ab} \phi_1^c
\xi_{-1}^c)\Gwh^{ab}\,,
\\[5pt]
\label{26112010-02} e^{-1}\delta_{b^a,\xi_\smminone}\LL_5 & = &
\phi_1^a(\phi_0^{ab}\xi_\smminone^b + \frac{1}{4}\phi_0^{bb}\xi_\smminone^ a
+ \frac{u}{2} \xi_\smminone^a \phi_0) \,.
\eeq
From these relations, we see that variation proportional to $\Gwh^{ab}$
\rf{2m18122010-06} can be cancelled by modifying $\xi_\smminone^a$ gauge transformation of the
field $\phi_2^{ab}$
\be  \label{2m18122010-07} \delta_{\xi_\smminone}'' \phi_2^{ab} = \phi_1^a
\xi_{-1}^b + \phi_1^b \xi_{-1}^a - \frac{1}{2} \eta^{ab} \phi_1^c \xi_{-1}^c
\,.\ee
Namely, it is easy to see that variation of $\LL_1$ \rf{22112010-01} under
gauge transformation of $\phi_2^{ab}$ \rf{2m18122010-07} cancels variation in
\rf{2m18122010-06}. Collecting results in \rf{2m18122010-03} and
\rf{2m18122010-07}, we find complete $\xi_\smminone^a$ gauge transformation
of the field $\phi_2^{ab}$,
\be \label{2m19122010-01} \delta_{\xi_\smminone} \phi_2^{ab} =
\LL_{\xi_\smminone^{\vphantom{}}}\phi_0^{ab} + \phi_1^a \xi_{-1}^b + \phi_1^b
\xi_{-1}^a - \frac{1}{2} \eta^{ab} \phi_1^c \xi_{-1}^c \,.\ee

{\bf Step 4}. Using notation $\delta_{\phi_2^{ab},\xi_\smminone}\LL_6$ for
variation of $\LL_6$ under $\xi_\smminone^a$ gauge transformation of field
$\phi_2^{ab}$ \rf{2m19122010-01}, we find
\be \delta_{\phi_2^{ab},\xi_\smminone}\LL_6 =
\delta_{\phi_2^{ab},\xi_\smminone}\LL_6|_{\phi_0^{ab}\phi_0^{ce},\phi_0^{ab}\phi_0}+
\delta_{\phi_2^{ab},\xi_\smminone}\LL_6|_{\phi_0^{ab}\phi_1^c,\phi_1^a\phi_0}\,,
\ee
\beq \label{2m19122010-02}
&&
e^{-1}\delta_{\phi_2^{ab},\xi_\smminone}\LL_6|_{\phi_0^{ab}\phi_0^{ab},\phi_0^{ab}\phi_0}
\equiv  - (\phi_0^2)^{ab}\DD^a \xi_\smminone^b + \phi_0^{ab}\phi_0^{cc}\DD^a
\xi_\smminone^b + \frac{1}{4} \phi_0^{ab}\phi_0^{ab}\DD\xi_\smminone
\nonumber\\[5pt]
&& \hspace{4cm}  -\frac{1}{4} \phi_0^{aa}\phi_0^{bb} \DD\xi_\smminone
+ \,\, \frac{u}{2}\phi_0(\xi\DD)\phi_0^{cc} + u
\phi_0^{ab}\phi_0 \DD^a\xi_\smminone^b \,,
\\[5pt]
\label{26112010-01} && e^{-1} \delta_{\phi_2^{ab},\xi_\smminone}
\LL_6|_{\phi_0^{ab}\phi_1^c,\phi_1^a\phi_0} \equiv  - \phi_1^a (
\phi_0^{ab}\xi_\smminone^b + \frac{1}{4}\phi_0^{bb}\xi_\smminone^a +
\frac{u}{2}\phi_0\xi_\smminone^a)\,.
\eeq
Comparing \rf{26112010-02} and \rf{26112010-01}, we find the cancellation
\be \delta_{b^a,\xi_\smminone}\LL_5 +
\delta_{\phi_2^{ab},\xi_\smminone}
\LL_6|_{\phi_0^{ab}\phi_1^c,\phi_1^a\phi_0}=0\,.
\ee
We proceed to the next step of our procedure with noticing that variations
that remain to be cancelled are given in
\rf{22112010-16a6},\rf{22112010-16a7}, and \rf{2m19122010-02}.

{\bf Step 5}.  We now consider $F^{ab}$ depending variation given in
\rf{22112010-16a6}. This variation can be cancelled by adding new
contributions to Lagrangian and modifying $\xi_\smminone^a$ gauge
transformations of the field $\phi_1^a$. Note that, in flat conformal
gravity, the field $\phi_1^a$ is not transformed under $\xi_\smminone^a$
gauge transformations (see \rf{2man29112010-02}). This is to say that, in
interacting conformal gravity, we consider the following new contributions to
Lagrangian and $\xi_\smminone^a$ gauge transformation of the field
$\phi_1^a$:
\beq
\label{25112010-01}
&& e^{-1}\LL_7 = c_1 F^{ac}F^{cb} \phi_0^{ab} + c_2 F^{ab}F^{ab} \phi_0^{cc}
+ c_3 F^{ab}F^{ab} \phi_0\,,
\\[5pt]
\label{25112010-03}
&&  \delta_{\xi_\smminone}\phi_1^a = f_1 \phi_0^{ab}\xi_\smminone^b + f_2
F^{ab}\xi_\smminone^b + f_3\phi_0 \xi_\smminone^a + f_4
\phi_0^{cc}\xi_\smminone^a\,,
\eeq
where coefficients $c_{1,2,3}$ and $f_{1,2,3,4}$ remain to be determined. To
this end we compute variations of $\LL_3$ \rf{22112010-03} and $\LL_5$
\rf{22112010-05} under $\xi_\smminone^a$ gauge transformation of the field
$\phi_1^a$ \rf{25112010-03},
\beq \label{25112010-04}
e^{-1}\delta_{\phi_1^a,\xi_\smminone}\LL_3 & = & f_2 \DD^a
F^{ab}F^{bc}\xi_\smminone^c
+   f_1 \DD^a F^{ab} \phi_0^{bc}\xi_\smminone^c
\nonumber\\[5pt]
& + & f_3 \DD^a F^{ab} \xi_\smminone^b\phi_0\,,
+  f_4 \DD^aF^{ab}\xi_\smminone^b \phi_0^{cc}\,,
\eeq

\be \label{25112010-05}
\delta_{\phi_1^a,\xi_\smminone}\LL_5 =
\delta_{\phi_1^a,\xi_\smminone}\LL_5|_F +
\delta_{\phi_1^a,\xi_\smminone}\LL_5|_{\phi_0^{ab},\phi_0}\,,
\ee
\beq \label{25112010-06}
e^{-1} \delta_{\phi_1^a,\xi_\smminone}\LL_5|_F & \equiv & f_2
F^{ab}\xi_\smminone^b (\DD\phi_0)^a
-  f_2 F^{ab}\xi_\smminone^b \DD^a\phi_0^{cc}
- f_2 u F^{ab}\xi_\smminone^b \DD^a\phi_0\,,
\eeq
\beq \label{25112010-06a1}
e^{-1}\delta_{\phi_1^a,\xi_\smminone}\LL_5|_{\phi_0^{ab},\phi_0} & \equiv &
f_1 \phi_0^{ab}\xi_\smminone^b (\DD^c\phi_0^{ac} - \DD^a\phi_0^{cc})
- f_1 u \phi_0^{ab}\xi_\smminone^b \DD^a\phi_0
\nonumber\\[5pt]
& + & f_3 \phi_0 \xi_\smminone^a (\DD\phi_0)^a
-  f_3 \phi_0 \xi_\smminone^a \DD^a\phi_0^{cc}
+  \half f_3 u \phi_0^2 \DD\xi_\smminone
\nonumber\\[5pt]
& + & f_4 \hbox{ terms}\,.
\eeq
Also, computing variation of $\LL_7$ \rf{25112010-01} under gauge
transformations \rf{22112010-16a1},\rf{22112010-16a2}, we find
\beq \label{25112010-07}
e^{-1}\delta_{\xi_\smminone} \LL_7 & = & -2 c_1 \DD^a F^{ab} F^{bc} \xi_\smminone^c
+  (2c_2 - \frac{c_1}{2}) F^2 \DD\xi_\smminone
-  2 c_1  F^{ac} \FF^{cb}(\xi_\smminone)\phi_0^{ab}
\nonumber\\[5pt]
& + & 4c_2 \DD^aF^{ab} \xi_\smminone^b \phi_0^{cc}
 + 4c_2 F^{ab} \xi_\smminone^b \DD^a \phi_0^{cc}
\nonumber\\[5pt]
& + &  4c_3  \DD^a F^{ab}\xi_\smminone^b \phi_0
+ 4c_3 F^{ab} \xi_\smminone^b \DD^a \phi_0\,.
\eeq
Requiring the $F^{ab}$ depending terms to cancel gives the equations
\be \label{25112010-09}
\delta_{\phi_0^{ab}, \xi_{-1}} \LL_2 \bigr|_F +
\delta_{\phi_1^a,\xi_\smminone}\LL_3 +
\delta_{\phi_1^a,\xi_\smminone}\LL_5|_F + \delta_{\xi_\smminone} \LL_7  = 0\,,
\ee
which allow us to fix all coefficients in \rf{25112010-01},\rf{25112010-03},
\be \label{25112010-10}
c_1 = - \frac{1}{4}\,, \quad c_2 = -\frac{1}{16}\,, \quad c_3 =- \frac{u}{8}
\,, \quad f_1 =-\half \,,\quad f_2 = -\half \,,\quad f_3 =
\frac{u}{2}\,,\quad f_4 = 0 \,.
\ee
Using \rf{25112010-10}, we note the relation $f_1u + f_3 = 0$, which allows
us to represent \rf{25112010-06a1} as (up to total derivative)
\beq \label{25112010-14}
e^{-1}\delta_{\phi_1^a,\xi_\smminone}\LL_5|_{\phi_0^{ab},\phi_0} & = & f_1
\phi_0^{ab}\xi_\smminone^b ( \DD^c\phi_0^{ac} -  \DD^a\phi_0^{cc})
 + f_1 u \phi_0^{ab} \phi_0 \DD^a \xi_\smminone^b
\nonumber\\[5pt]
& - & f_3 \phi_0 \xi_\smminone^a \DD^a\phi_0^{cc}
+ \half f_3 u \phi_0^2 \DD\xi_\smminone\,.
\eeq

{\bf Step 6}. Variations that remain to be cancelled are given in
\rf{22112010-16a7},\rf{2m19122010-02},\rf{25112010-14}. All these variations
involve terms of the second order in the fields $\phi_0^{ab}$ and $\phi_0$.
Note that the variation of $\LL_4$ \rf{22112010-04} under $\xi_\smminone^a$
gauge transformation also gives terms of the second order in the field
$\phi_0$,
\be \label{2m19122010-04} e^{-1} \delta_{\xi_\smminone} \LL_4 = -
\frac{1}{4}\phi_0^2 \DD^c\xi_\smminone^c \,. \ee
We note that variations
\rf{22112010-16a7},\rf{2m19122010-02},\rf{25112010-14} and \rf{2m19122010-04}
can be cancelled by adding new contributions to Lagrangian without any
additional modification of $\xi_\smminone^a$ gauge transformations of the fields.
This is to say that we consider the following new contributions to
Lagrangian:
\beq \label{2m19122010-05}
e^{-1} \LL_8 & = & p_1\phi_0^{ab}\phi_0^{bc}\phi_0^{ca} +
p_2\phi_0^{ab}\phi_0^{ab}\phi_0^{cc} + p_3(\phi_0^{aa})^3
\nonumber\\[5pt]
& + & p_4\phi_0^{ab}\phi_0^{ab}\phi_0 + p_5 \phi_0^{aa}\phi_0^2 + p_6\phi_0^3
+ p_7(\phi_0^{aa})^2 \phi_0\,.
\eeq
Computing
\beq
&& e^{-1} \delta_{\xi_\smminone} \LL_8 = 6p_1 (\phi_0^2)^{ab}\DD^a\xi_\smminone^b +
4p_2\phi_0^{ab}\phi_0^{cc}\DD^a\xi_\smminone^b +
2p_2(\phi_0^2)^{aa}\DD\xi_\smminone + 6p_3\phi_0^{aa}\phi_0^{bb}
\DD\xi_\smminone \qquad
\nonumber\\[5pt]
&&\hspace{2cm}  +  \,\, 4p_4 \phi_0^{ab}\phi_0\DD^a\xi_\smminone^b +
2p_5\phi_0^2\DD\xi_\smminone + 4p_7 \phi_0^{aa} \phi_0 \DD\xi_\smminone\,,
\eeq
and requiring
\be \delta_{b^a,\xi_{-1}} \LL_2 + \delta_{\xi_\smminone} \LL_4 +
\delta_{\phi_1^a,\xi_\smminone}\LL_5|_{\phi_0^{ab},\phi_0} +
\delta_{\phi_2^{ab},\xi_\smminone}\LL_6|_{\phi_0^{ab}\phi_0^{ce},\phi_0^{ab}\phi_0}
+ \delta_{\xi_\smminone} \LL_8 = 0 \,, \ee
we get
\beq
&& p_1 = \frac{1}{4}\,,\quad \ p_2 = -\frac{5}{16}\,,\quad \ p_3 =
\frac{1}{16}\,, \quad \ p_4 = -\frac{u}{8}\,, \quad \ p_5 = -\frac{3}{16}\,,
\quad \ p_7 = 0\,.\qquad
\eeq
Thus we see that, with exception of the coefficient $p_6$,  all the remaining
coefficients entering cubic potential \rf{2m19122010-05} are fixed by
$\xi_\smminone^a$ gauge symmetries. Note also that all variations of the
Lagrangian $\LL$, which are proportional to the gauge transformation
parameter $\xi_\smminone^a$, have been cancelled.

{\bf Step 7}.  The coefficient $p_6$ is fixed by considering $\xi_0$ gauge
transformations.  With exception of the field $\phi_2^{ab}$, $\xi_0$ gauge
transformations given in \rf{2m01122010-34}-\rf{2m01122010-39} are obtained
by covariantization of the corresponding gauge transformations of free
fields. Using gauge transformations in \rf{2m01122010-34}-\rf{2m01122010-39}
we check the relations
\be
\delta_{\xi_0} ( \LL_{2\smK} + \LL_4 + \LL_5 )  = 0\,,\qquad \delta_{\xi_0} (
\LL_3 + \LL_7 ) = 0\,,
\ee
where we use the decomposition $\LL_2 = \LL_{2\smK} + \LL_{2\smR}$ (see
\rf{2m01122010-43},\rf{2m18122010-01},\rf{2m18122010-02}). We note that
$\LL_{2\smR}$ \rf{2m18122010-02} is not invariant under $\xi_0$ gauge
transformation
\be e^{-1}\delta_{\xi_0}\LL_{2\smR} = -\half \phi_0^{ab}\Gwh^{ab}\xi_0 \,.\ee
It is easy to see that this gauge variation can be cancelled by modifying
$\xi_0$ gauge transformation of the field $\phi_2^{ab}$,
\be \delta_{\xi_0}'\phi_2^{ab} = -\half \phi_0^{ab}\xi_0\,.\ee
We now consider the remaining gauge variations to be cancelled
\beq
&&  e^{-1}\delta_{\xi_0}\LL_8 =  -\frac{1}{4}\phi_0^{ab}\phi_0^{ab}\xi_0 +
\frac{1}{4}\phi_0^{aa}\phi_0^{bb}\xi_0 + \frac{u}{4}\phi_0^{aa}\phi_0\xi_0  -
3(\frac{3}{16} +  up_6)\phi_0^2\xi_0\,,
\\[5pt]
&& e^{-1}\delta_{\xi_0}' \LL_6 = \ \ \frac{1}{4}\phi_0^{ab}\phi_0^{ab}\xi_0 -
\frac{1}{4}\phi_0^{aa}\phi_0^{bb}\xi_0 - \frac{u}{4}\phi_0^{aa}\phi_0\xi_0\,.
\eeq
Requiring $\delta_{\xi_0}' \LL_6 + \delta_{\xi_0} \LL_8 = 0$, we get $p_6
=-\frac{3}{16u}$.

Thus, with exception of $\xi_1^a$ gauge transformations, we have checked
gauge invariance of our Lagrangian with respect to all gauge transformations.
The $\xi_1^a$ gauge transformations of interacting theory
\rf{2m01122010-22}-\rf{2m01122010-27} are simply obtained by
covariantization, $\partial^a\rightarrow \DD^a$, of the ones of flat theory
\rf{2m01122010-05}-\rf{2man29112010-03}. Doing so, we note that only the
contributions to Lagrangian denoted by $\LL_1$, $\LL_3$, $\LL_5$, $\LL_6$ are
changed under $\xi_1^a$ gauge transformations. Using the easily derived
relations
\be
\delta_{\xi_1} ( \LL_1   + \LL_3 )  = 0\,, \qquad  \delta_{\xi_1} ( \LL_5  + \LL_6 ) = 0 \,,
\ee
we see that Lagrangian \rf{2m04122010-01} is invariant under the $\xi_1^a$ gauge
symmetries. This finishes our procedure of building the gauge invariant
Lagrangian and the corresponding gauge transformations.

\end{document}